\documentclass[a4paper, fleqn]{cas-sc}
\usepackage{courier}  
\usepackage{graphicx} 
\usepackage[authoryear,longnamesfirst]{natbib}  
\usepackage{caption} 
\usepackage{algorithm}
\usepackage{algorithmic}
\usepackage{amsmath} 
\usepackage{newfloat}
\usepackage{listings}
\usepackage{enumitem}
\usepackage{amssymb}
\usepackage{booktabs}
\usepackage{subcaption}
\usepackage{multirow}
\usepackage{pifont}

\usepackage{cite}
\usepackage{amsfonts}
\usepackage{textcomp}
\usepackage{xcolor}
\usepackage{array} 
\usepackage{makecell}
\usepackage{hyperref}  

\usepackage{threeparttable} 
\usepackage{xcolor}

\definecolor{customblue}{RGB}{58, 134, 255}



\begin{document}
\let\WriteBookmarks\relax
\def\floatpagepagefraction{1}
\def\textpagefraction{.001}

\author[1]{Jiale Xu}[style=chinese]
\ead{x17329348016@email.swu.edu.cn}

\author[1]{Mingshu Ai}[style=chinese]
\ead{amssma9@email.swu.edu.cn}

\author[1]{Junjie Huang}[style=chinese]
\corref{cor}
\ead{junjiehuang@swu.edu.cn}

\author[1,2]{Tao Jia}[style=chinese]
\corref{cor}
\ead{tjia@swu.edu.cn}
\cortext[cor]{Corresponding author}

\affiliation[1]{organization={College of Computer and Information Science, Southwest University},
    city={Chongqing},
    postcode={400715}, 
    country={China}}

\affiliation[2]{organization={College of Computer and Information Science, Chongqing Normal University},
    city={Chongqing},
    postcode={401331}, 
    country={China}}

\shorttitle{Path-Augmenting Graph Structure Learning for explainable scholar recommendation}
\title [mode = title]{PAGSL: Path-Augmenting Graph Structure Learning for explainable scholar recommendation}

\begin{abstract}
Explainable scholar recommendation aims to identify potential scholars of interest and provide reasonable explanations for the recommendations. However, existing frameworks struggle with optimization-induced noise and "hub node bias", yielding trivial explanations that lack hierarchical depth and personalization.
To this end, we propose a novel explainable scholar recommendation model, Path-Augmenting Graph Structure Learning (named PAGSL). 
PAGSL combines representation learning and semantic path knowledge to ensure recommendation performance while providing interpretable results from two perspectives. 
Specifically, it activates latent academic paths among entities through graph augmentation and employs a path-based neural network to encode these paths for recommendation. 
A masking mechanism with path score optimization is designed to generate concise and informative explanations. 
We evaluate PAGSL on three academic datasets (Scholat, AMiner, and ResearchGate), comprising approximately 49,000 authors and over 330,000 interaction records. Experimental evaluations validate that PAGSL outperforms state-of-the-art baselines, achieving significant improvements of 6.31\% in F1-score, 6.29\% in Accuracy and 5.95\% in AUC for recommendation. Crucially, in terms of explainability, we provide quantitative results and visualization effects. Furthermore, ablation studies validate the effectiveness of our heterogeneous convolution and path representation learning. PAGSL achieves strong performance in both recommendation accuracy and explainability.

\end{abstract}


\maketitle

\section{Introduction}
Driven by the exponential expansion of research output, the academic community has transitioned into an era of data-driven exploration. In this context, identifying potential collaborators and monitoring emerging research trends pose a significant challenge for researchers. Consequently, scholar recommendation systems based on Academic Social Networks (ASNs) have become increasingly important in real-world scenarios, supporting tasks such as reviewer assignment and collaborator discovery~\citep{aksoy2023reviewer,TEDRec2025}. Unlike conventional e-commerce recommendations, scholar recommendations operate within a stringent, trust-centric decision-making environment. Opaque outcomes, despite high accuracy, are often dismissed by experts if they cannot provide rationales grounded in academic principles~\citep{ZhangShichang2023,LiYicong2024}. Thus, integrating predictive precision with explainability remains a fundamental challenge in ASNs analysis.

Lately, Graph Neural Networks (GNNs) have emerged as the mainstream paradigm for modeling ASNs due to their remarkable capability for processing non-Euclidean data. By employing message passing and neighborhood aggregation over heterogeneous graphs, GNNs can effectively capture high-order dependencies between scholarly entities~\citep{CNEacR2021,HNERec2023}. More recently, some research incorporates contrastive learning to enhance model performance in different scenarios~\citep{HGCL2023,Pone-GNN2025,SDCGCL2025}. However, existing GNN-driven scholar recommendation methods still face two critical limitations in practical applications.

\textbf{The sparsity and noise of ASNs impose fundamental constraints on representation learning quality}.  Conventional interpretable models ~\citep{PGExplainer2020,GNNExplainer2019,SubgraphX2021} predominantly employ subgraph-based reasoning over observed co-occurrence networks, such as co-authorship graphs. However, these academic networks often manifest severe cold-start problems, and their explicit edges frequently fail to capture latent semantic dependencies. As a result, researchers with highly similar interests may remain topologically disjoint due to a lack of intermediary connections, preventing models from identifying collaborative opportunities driven by attribute-level similarity rather than mere structural proximity. Although meta-path-based methods~\citep{TMER2021,ZhengJiayin2022,CaGE2025} attempt to address these limitations, they are often constrained by their reliance on predefined heuristics and their inability to adapt to the dynamic evolution of graph structures, ultimately undermining recommendation robustness.

\begin{figure}
	\centering
		\includegraphics[scale=.5]{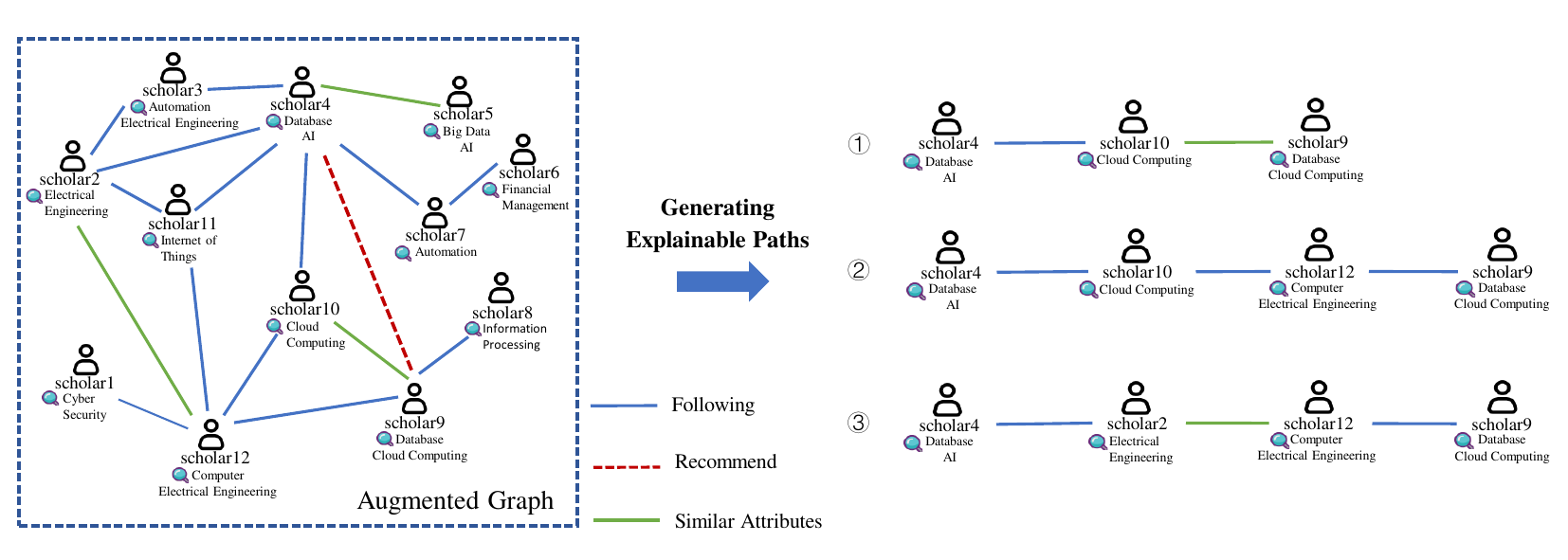}
	\caption{A case study for generating explainable paths. For `scholar 4' and `scholar 9', path `\ding{172}-\ding{174}' denote the available paths between them. Path `\ding{172}' is an ideal explanation, path `\ding{173}' and `\ding{174}' are the less informative explanations.}
    \vspace{-0.3cm}
    \label{fig:fig-1}
\end{figure}

\textbf{Explanation path generation is frequently compromised by the "hub node bias" inherent in network topologies}. Prevailing path generation paradigms typically adopt path walks~\citep{GengShijie2022,YangCao2024} and attention mechanisms~\citep{TGAMER2023,CPER2024,KPAR2025} to find the shortest or high-affinity trajectories. In scale-free academic networks, this leads to an inevitable convergence of explanation paths on a limited set of high-impact hub nodes, such as prominent leading scholars. Consequently, this phenomenon yields homogenized and trivialized explanations that frequently cite generic connections to renowned figures while neglecting the local core structures that define specific research subfields. 
As shown in Figure \ref{fig:fig-1}, path `\ding{172}' is a meaningful explanation. `Scholar 4' follows `Scholar 10', then `Scholar 10' and `Scholar 9' have similar interests and preferences. Therefore, `Scholar 4' may also be interested in `Scholar 9'. Conversely, paths `\ding{173}' and path `\ding{174}' traverse prominent hubs like `Scholar 12', whose high degree reflects a generalized presence across the `Computer' domain rather than specialized expertise. Attention-based methods may assign higher weights to such paths, and fail to capture paths with specific explanatory semantics such as `Cyber Security'.
Such a deficiency in hierarchical structure awareness results in explanations that, despite maintaining topological connectivity, are considerably diminished in semantic depth and personalization.

To address these limitations, we propose a novel framework named \textbf{P}ath-\textbf{A}ugmenting \textbf{G}raph \textbf{S}tructure \textbf{L}earning (\textbf{PAGSL}) for explainable scholar recommendation. 
PAGSL simultaneously refines noisy graph structures and models semantically meaningful relational paths, enabling more accurate and interpretable recommendations.
PAGSL consists of three key components. First, an attribute-driven augmentation strategy enriches the academic social graph by incorporating edges based on shared scholar attributes, which activates potential semantic paths. Second, a multi-relational GNN is developed to encode diverse interaction paths among scholars. Each scholar’s representation is updated by aggregating multiple behavioral paths, supporting robust and semantically grounded recommendation. Third, an edge-mask learning mechanism identifies influential paths, while a path evaluation function selects concise and informative relational chains as final explanations.
We evaluate PAGSL on three real-world academic datasets using multiple recommendation and explanation metrics. The empirical evidence underscores that PAGSL consistently exceeds strong baselines in both recommendation performance and explainability.

Our main contributions are summarized as follows:
\vspace{-5pt}

\begin{itemize}
    \item We propose a novel explainable scholar recommendation framework, \textbf{PAGSL}, which unifies graph structure refinement with interpretable semantic path modeling, effectively addressing both noise reduction and user-centric explanation in academic social networks.
    
    \item We develop a path-aware relation learning module that integrates augmented path trees with heterogeneous encoding. By employing an attribute-sharing augmentation strategy, we establish semantic bridges to improve subsequent path reasoning and resolve reachability issues for cold-start nodes. This path trees-based heterogeneous encoding architecture preserves the branching propagation of academic influence, overcoming the expressive bottlenecks of sequential paths in capturing complex contextual nuances.

    \item We design a topological path explanation strategy based on hierarchical decomposition guided mask learning and path score optimization. Our module prioritizes robust paths that traverse the core layers of the network, effectively filtering out noise and trivial hub-node transitions. This enables the generation of explanatory paths characterized by high fidelity and structural significance, thereby achieving dual improvement in both recommendation precision and explainability.
    
    \item Through experimental evaluations on three real-world academic social networks, the results reveal that PAGSL surpasses existing state-of-the-art methods, effectively bridging the gap between scholar recommendation effectiveness and explanatory quality.
\end{itemize}

\section{Related Work}
\label{sec:related_work}

\subsection{Scholar Recommendation}
Scholar recommendation aims to identify relevant scholars for a given user based on academic interests, social interactions, and attribute signals.
Current research on scholar recommendation includes collaborative filtering methods and deep learning methods. 
For example, NNRW~\citep{NNRW2018} integrates a nearest neighbor filtering with a PageRank to mitigate the data sparsity issue. HeteroRWR~\citep{HeteroRWR2020} integrates heterogeneous random walks and multi-relational weight adaptation for interdisciplinary collaborator recommendations. 
Deep learning methods enhance the modeling of linear and nonlinear interactions through GNNs. For example, CNEacR~\citep{CNEacR2021} constructs a textual similarity measurement module based multi-head attention and a dynamic fusion strategy. HNERec~\citep{HNERec2023} designs a hierarchical graph attention and a semantic fusion coefficients for improving long-tail scholar matching. 
To enhance robustness in sparse or noisy scenarios, recent advancements have incorporated contrastive learning and self-supervised learning paradigms into heterogeneous graphs. For example, SGL~\citep{SGL2021} constructs multi-view graph structural augmentation strategies and a contrastive learning mechanism. HGCL~\citep{HGCL2023} encodes the heterogeneous user-item interaction and integrates a meta network with contrastive learning to advance the recommendation.
To further capture rich semantic relations, HSGL~\citep{HSGL2025} introduces a heterogeneous self-supervised graph learning framework that maps relation-specific embeddings into a unified space via cross-relation contrastive learning. 
However, these methods often neglect explainability and fine-grained semantic paths essential for transparent and user-trustworthy recommendations.

GNN-based models have shown strong capabilities in capturing complex relations in ASNs. 
Nevertheless, their effectiveness is susceptible to the integrity of the input graph topology, which is often noisy, sparse, or incomplete in real-world ASNs~\citep{PTDNet2021}. 
To address this, Graph Structure Learning (GSL) has been proposed to jointly optimize graph topology and node representations. SLHRec~\citep{SLHRec2024} jointly models heterogeneity and logical dependencies by collaborating on a structure-aware dual-GNN with a logic-aware Markov network through a contrastive objective. GCEN~\citep{GCEN2024} formulates a self-supervised contrastive framework to refine graph connectivity via integrating multi-branch generative modules with feature enhancement. DSL-MRS~\citep{DSL-MRS2025} enhances recommendation through a structured diffusion process that counteracts sparsification-induced information loss by preserving global structural integrity. Furthermore, to address the temporal nature of interactions, DGEL~\citep{DGEL2023} explicitly models how node neighborhoods and structural contexts evolve over time to capture changing user preferences.

Similarly to these works, our PAGSL framework adopts the structural refinement to combat sparsity and improve recommendation quality. However, our attribute-sharing augmentation strategy provides richer signals for the path encoding module and creates explicit semantic bridges that form the basis for interpretable path-based explanations. This effectively resolve the reachability problem for cold-start scholars. 
Furthermore, rather than flattening interactions into sequential linear paths, our path-tree-based architecture strictly preserves the branching propagation of academic information. This overcomes the expressive limitations of linear paths while simultaneously providing a transparent, tree-structured topological foundation that directly facilitates our end-to-end explainable recommendation.

\subsection{Explainable Recommendation}
Explainable recommendation seeks to provide human-understandable rationales for prediction outcomes. Subgraph-based methods focus on identifying local substructures important to prediction. 
For example, GNNExplainer~\citep{GNNExplainer2019} learns parameter-driven masks to extract edge-induced subgraphs and critical features. Its objective is to maximize the mutual information between the masked structure and the original predictive output. PGEXplainer~\citep{PGExplainer2020} uses the same mutual information, but employs a neural mask generator to facilitate the production of explanatory structures. However, they often fail to capture long-range semantic dependencies.
Path-based methods provide more intuitive explanations by tracing relational chains between users and items. For example, ERKM~\citep{ERKM2021} generates the path between entities as an explanation basis, and optimizes the explanation list based on multi-objective algorithm. CaGE~\citep{CaGE2025} integrates structural causal models to mitigate confounding bias and generates meta-path explanations that highlight true causal pathways.
However, they suffer from redundancy and often rely on domain-specific meta-paths~\citep{MEIRec2019,TMER2021}, limiting generalizability.

Attention-based methods learn importance weights over multiple paths based on attention mechanism. For example,  TGAMER~\citep{TGAMER2023} utilizes proximity-aware selection to retrieve meta-path samples that contribute significantly to the model, and processes path information based on an attention mechanism. TMER-RL~\citep{TMER-RL2023} generates useful instance paths based on reinforcement learning, and learns the weights of instances that contain reasoning information with multi-head attention. DCCMA-Net~\citep{DCCMA-Net2025} separates features into content and style components to align semantic correlations weight between different entities.

However, these methods often suffer from the hub-node bias inherent in scale-free academic networks, frequently yielding generic, uninformative explanations because they fail to disentangle topology from popularity.
Like subgraph-based methods, we employ learnable edge masks to identify task-relevant structures in a differentiable manner. Like path-based and attention-based methods, our final explanations take the form of concrete relational paths between scholars and we also evaluate path importance. 
However, our differentiation lies in actively intervening in the graph topology. By integrating a $k$-shell decomposition to eliminate topological redundancy, and employing a centrality penalty within the mask learning objective, our framework explicitly suppresses the interference of generic hub nodes. Consequently, PAGSL does not require manual path filtering or domain-specific meta-paths; it automatically navigates the informative middle shells of the network to output concise, specific, and semantically meaningful reasoning paths.

\section{Problem Definition}

We address the task of explainable scholar recommendation over Academic Social Networks (ASNs), where the goal is twofold: (1) to recommend  relevant scholars to a given user, and (2) to provide personalized, explainable justifications for the recommendation.

Formally, we define an ASN as a directed attributed graph $G = (V, E, A)$, where $V$ denotes scholar nodes, $E$ encodes academic relations (e.g., co-authorship, citation), and $A$ captures node-level attributes such as research interests. Given a target scholar $u \in V$, the recommendation objective is to identify
\begin{equation}
v^* = \arg\max_{v \in V} \phi(u, v),
\end{equation}
where $\phi(u,v)$ estimates the scholarly relevance between $u$ and $v$. Additionally, the model should produce a semantic path $P_{u \rightarrow v^*}$ that explains the recommendation based on structural evidence in $G$.

\section{Proposed Model}

\label{sec:Proposed Model}
\begin{figure*}[t]
    \centering
    \includegraphics[width=1.0\linewidth]{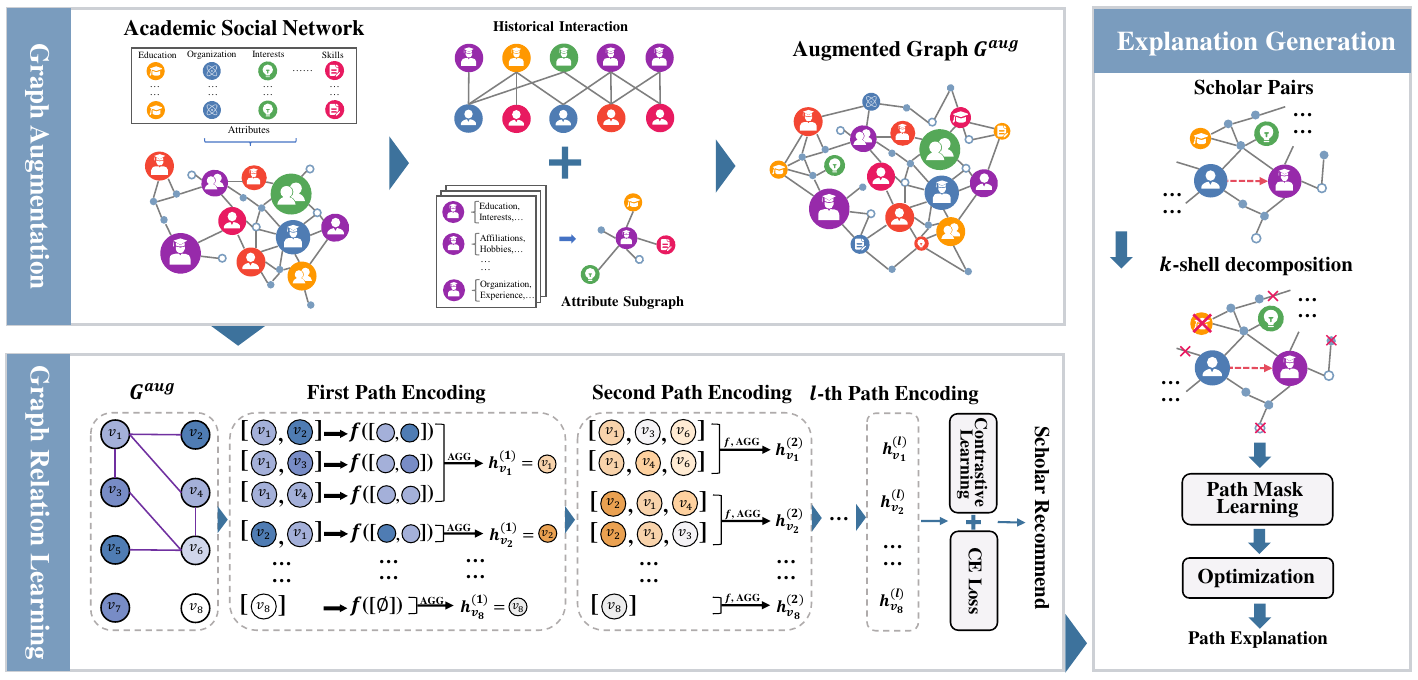}
    \caption{The Framework of the PAGSL framework. PAGSL includes three key modules: (1) Graph augmentation for activating semantic paths based on attribute subgraph relationship. (2) Relation paths extraction and representation in augmented graph for scholar recommendation by path trees construction and path neural network. (3) Concise and informative path explanations generation with path mask learning and explanation optimization.}
    \label{fig:framework}
\end{figure*}

In this section, we introduce our  Path-Augmenting Graph Structure Learning (PAGSL) model, which consist of \textit{Graph Augmentation}, \textit{Graph Relation Learning}, and \textit{Explanation Generation}.
The overall architecture of PAGSL is illustrated in Figure~\ref{fig:framework}.

\subsection{Graph Augmentation}
\label{sec:G_aug}
The scholars' attributes in ASNs are generally represented as vectors. Although this can improve recommendation performance, it also reduces available information during result explanation.
To address the issues of sparse connectivity and the lack of explicit semantic edges~\citep{Scholar2vec2021,SI-HDGNN2022,HNERec2023}, we introduce a graph augmentation strategy that leverages attribute-level connections to enrich the original network structure. 
Specifically, we transform scholar attributes into heterogeneous edges, enabling the model to capture more semantic relations and activate additional explanatory paths between users and scholars, as illustrated in Figure~\ref{fig:framework}.

Concretely, for each scholar node $v_i$, we first construct an attribute subgraph $G^{sub}_{v_i}$ that connects $v_i$ to its associated attribute nodes via bidirectional edges:
\begin{equation}
    G^{sub}_{v_i} = \left( N_{v_i}, E_{v_i} \right),
\end{equation}
where $N_{v_i}$ denotes the attribute nodes (e.g., research fields, affiliation), and $E_{v_i}$ represents the set of edges linking $v_i$ to its attributes.

Next, we compare the subgraphs across scholars to identify those sharing common attributes and add type-specific edges between them. This produces an augmented graph:
\begin{equation}
    G^{aug} = \left( V, E^{aug} \right),
\end{equation}
where $V$ is the set of scholar nodes and $E^{aug}$ includes both original follow edges and newly added attribute-sharing edges. 
Finally, we extract all candidate paths between user–scholar pairs from the augmented graph $G^{aug}$.

\subsection{Graph Relation Learning}
\begin{figure*}
    \centering
    \includegraphics[width=\linewidth]{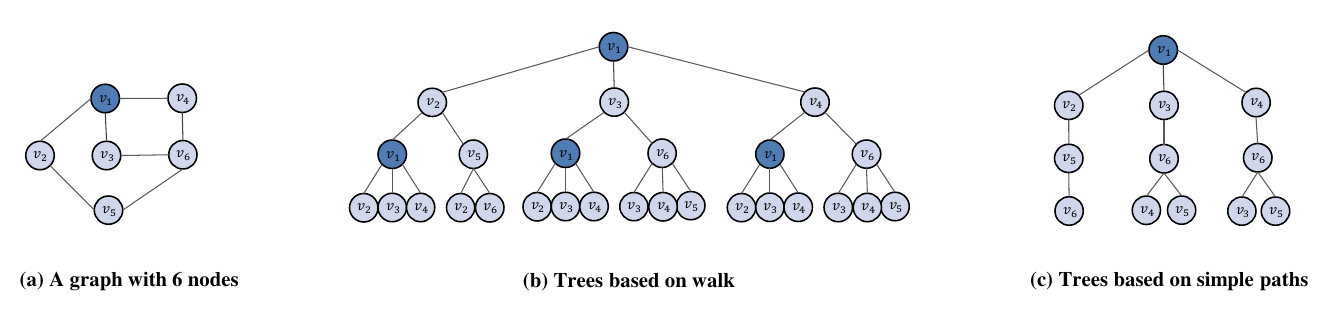}
    \caption{Illustration of constructing behavioral path trees. (a): Example graph with 6 nodes. (b): Tree of depth 3 generated via random walk from $v_1$. This structure inherently allows repeated node visits(e.g., $v_1 \rightarrow v_2 \rightarrow v_1$), which bounds its expressiveness to the 1-WL test. (c): Tree built using all simple paths from $v_1$. By explicitly forbidding cycles, this structure effectively truncates structural redundancy and preserves semantic integrity.}
    \label{fig:path-generate}
\end{figure*}

In this subsection, we design a relation learning module that transforms heterogeneous academic interactions into effective node representations for personalized scholar recommendation.
To model heterogeneous behavior paths, we organize each scholar's local graph neighborhood as a behavioral path tree $\mathcal{T}_{v_i}$ rooted at node $v_i$, enabling structured aggregation over multiple interaction trajectories. 
Figure~\ref{fig:path-generate} illustrates two strategies for constructing behavioral path trees. 

Formally, given $G^{aug} = (V, E^{aug})$, we construct a tree $\mathcal{T}_{v_i}^{d}$ rooted at scholar $v_i$ by enumerating all simple paths from $v_i$ up to a maximum depth $d$. 
Each node at layer $l$ represents a scholar or attribute entity reached via a simple path of length $l$, and it connects only to nodes at layer $l+1$ that appear in the next hop of the path. 
Random walks trees $\mathcal{T}_{rw}^d(v_i)$ (Figure~\ref{fig:path-generate}b) enumerate all walks of length up to $d$ from node $v_i$, allowing repeated node visits. This process is structurally equivalent to the message-passing GNNs and is bounded by the expressive power of the 1-dimensional Weisfeiler-Lehman (1-WL) test. In contrast, the simple-path trees $\mathcal{T}_{sp}^d(v_i)$ (Figure~\ref{fig:path-generate}c) effectively truncates this redundancy by explicitly forbidding cycles and better preserves semantic integrity. We formalize its superiority over the random walk tree in the following proposition.

Proposition 1 (Strict Expressiveness Hierarchy \citep{PathNN2023}). Let $G = (V, E)$ be a graph. For any depth $d$ and any two nodes $v_i, v_j \in V$. 
(1) If $\mathcal{T}_{rw}^d(v_i) \not\simeq \mathcal{T}_{rw}^d(v_j)$ (i.e., 1-WL distinguishes $v_i$ from $v_j$), then $\mathcal{T}_{sp}^d(v_i) \not\simeq \mathcal{T}_{sp}^d(v_j)$.
(2) There exist graphs where $\mathcal{T}_{rw}^d(v_i) \simeq \mathcal{T}_{rw}^d(v_j)$ for all $d$, yet $\mathcal{T}_{sp}^d(v_i) \not\simeq \mathcal{T}_{sp}^d(v_j)$ for some finite $d$.

Proof. (1) We prove the contrapositive $\mathcal{T}_{sp}^d(v_i) \simeq \mathcal{T}_{sp}^d(v_j) \Rightarrow \mathcal{T}_{rw}^d(v_i) \simeq \mathcal{T}_{rw}^d(v_j)$ by induction on $d$. For $d > 1$, assume the claim holds at $d-1$. The isomorphism $\mathcal{T}_{sp}^d(v_i) \simeq \mathcal{T}_{sp}^d(v_j)$ induces a bijection $\pi: \mathcal{N}(v_i) \to \mathcal{N}(v_j)$ between neighbors. For any $u \in \mathcal{N}(v_i)$, its unconditional simple-path tree $\mathcal{T}_{sp}^{d-1}(u)$ consists of paths moving away from $v_i$ and paths routing back through $v_i$. Because the entire depth-$d$ trees of $v_i$ and $v_j$ are exactly isomorphic, both topological directions are preserved, guaranteeing $\mathcal{T}_{sp}^{d-1}(u) \simeq \mathcal{T}_{sp}^{d-1}(\pi(u))$. By the inductive hypothesis, this implies $\mathcal{T}_{rw}^{d-1}(u) \simeq \mathcal{T}_{rw}^{d-1}(\pi(u))$. Consequently, under the 1-WL algorithm, these paired neighbors will receive identical representation labels at iteration $d-1$. Since $v_i$ and $v_j$ aggregate from identical multisets of neighbor labels, they are guaranteed to receive the same updated label at iteration $d$, thereby establishing $\mathcal{T}_{rw}^d(v_i) \simeq \mathcal{T}_{rw}^d(v_j)$.
(2) Let $G = C_3 \sqcup C_6$ (the disjoint union of a triangle and a hexagon), with $v_i \in C_3$ and $v_j \in C_6$. Both $C_3$ and $C_6$ are 2-regular, so every node has the same initial state and every neighbor multiset is identical at each iteration. Thus, 1-WL never distinguishes any two nodes, yielding $\mathcal{T}_{rw}^d(v_i) \simeq \mathcal{T}_{rw}^d(v_j)$ for all $d$. However, simple-path trees differ structurally. In $C_3$, every simple path from $v_i$ visits at most 3 nodes, so $\mathcal{T}_{sp}(v_i)$ has a maximum depth of 2. In $C_6$, simple paths from $v_j$ can extend to all 6 nodes, yielding a maximum depth of 5. At $d = 3$, $\mathcal{T}_{sp}^3(v_i)$ has no nodes at depth 3 while $\mathcal{T}_{sp}^3(v_j)$ does. Therefore, $\mathcal{T}_{sp}^3(v_i) \not\simeq \mathcal{T}_{sp}^3(v_j)$.

In the context of our academic recommendation framework, this theoretical proof guarantees that simple-path trees are not merely a heuristic to reduce computational cost. Instead, they empower the model to capture deep and distinctive structural semantics (such as diverse collaborative circles).
This structure serves as the candidate path space for the downstream masking mechanism and explanation generation module, providing a transparent and explainable set of relational paths. This design avoids cyclic walks and enhances explainability by focusing on non-redundant relational paths.

Let $\mathcal{P}_{v_i}$ denote the set of all simple paths rooted at $v_i$, where each path $\varphi = \{v_0, v_1, \ldots, v_l\}$ satisfies $v_0 = v_i$ and contains no repeated nodes. 
Then the $l$-th layer of the path tree is:
\begin{equation}
    \mathcal{T}^{(l)}_{v_i} = \{ v_j \mid \exists~ \varphi \in \mathcal{P}_{v_i},~ \varphi(l) = v_j \},
\end{equation}
and two nodes $v_m \in \mathcal{T}^{(l)}_{v_i}$ and $v_n \in \mathcal{T}^{(l+1)}_{v_i}$ are connected iff $\exists~ \varphi \in \mathcal{P}_{v_i}$ such that $\varphi(l) = v_m$ and $\varphi(l+1) = v_n$.

\noindent\textbf{Path encoding.} To encode paths of length $l$, we apply a path encoder function $f(\cdot)$ (e.g., LSTM~\citep{LSTM1997}) over the sequence of node embeddings from the previous layer:
\begin{equation}
    h_p^{(l)} = f\left([h_{v_l}^{(l-1)}, h_{v_{l-1}}^{(l-1)}, \ldots, h_{v_0}^{(l-1)}] \right),
\end{equation}
where path $p = \left\{ {v_{0},v_{1},\ldots,v_{l}} \right\}$ denotes a behavioral path of length $l$ from source node $v_{0}$ to target node $v_{l}$.

\noindent\textbf{Heterogeneous path aggregation.} Inspired by R-GCN~\citep{RGCN2018}, we aggregate paths by relation type. Let $R$ be the set of relation types, and $P_{v_i}^{(l-1)}(r)$ be the set of length-$(l-1)$ paths of type $r$ starting from $v_i$. The updated representation of node $v_i$ at layer $l$ is:

\begin{equation}
h_{v_i}^{(l)} = \text{ReLU}\big(
    \sum_{r \in R} \frac{1}{|P_{v_i}^{(l-1)}(r)|} \sum_{p_j \in P_{v_i}^{(l-1)}(r)} W_r^{(l-1)} h_{p_j}^{(l-1)} + W_0^{(l-1)} h_{v_i}^{(l-1)} \big)
\end{equation}
where $W_r^{(l-1)}$ and $W_0^{(l-1)}$ are trainable weights, and the relation type $r$ is determined by the edge type between the center node and its direct neighbor.

\noindent\textbf{Prediction.} After $d$ layers of path encoding and aggregation, we obtain the final representation $h_{v_i}^{(d)}$ for each scholar. A prediction head is used to compute the matching score between a user $u$ and a candidate scholar $v$.

\noindent\textbf{Training objective.}
The model is jointly optimized through a composite objective encompassing both supervised recommendation loss and contrastive learning loss.

\textit{(i) Recommendation loss.}
We adopt a binary cross-entropy loss over observed and negatively sampled user-scholar pairs:
\begin{equation}
\begin{split}
    \mathcal{L}_{ce} = -\frac{1}{|E \cup E^{-}|} \sum_{(v_i, v_j) \in E \cup E^-}
    \left[
        y_{ij} \log \sigma(\hat{y}_{ij}) + 
        (1 - y_{ij}) \log (1 - \sigma(\hat{y}_{ij}))
    \right],
\end{split}
\end{equation}
where $\hat{y}_{ij}$ is the predicted score, $y_{ij}$ is the ground truth label, and $\sigma(\cdot)$ is the sigmoid function.

\textit{(ii) Contrastive loss.}
To enhance representation quality, we incorporate an InfoNCE contrastive loss~\citep{chen2020simple}:
\begin{equation}
    \mathcal{L}_{cl} = -\sum_{(v_i, v_j) \in E}
    \log \frac{
        F(v_i,v_j)
    }{
        F(v_i,v_j) + \sum_{(v_i, v_j') \in E^-}
        F(v_i,v_j')
    },
\end{equation}
where $F(v_i, v_j)=\exp(s(h_{v_i}, h_{v_j}) / \tau)$ is the match degree between sample pairs, $s(\cdot, \cdot)$ denotes cosine similarity and $\tau$ is the temperature parameter.

\textit{(iii) Overall loss.}
The final objective combines both components:
\begin{equation}
    \mathcal{L}_{rec} = \mathcal{L}_{ce} + \lambda_{\text{InfoNCE}} * \mathcal{L}_{cl},
\end{equation}
where $\lambda_{\text{InfoNCE}}$ balances the impact of contrastive learning. 
The pseudocode of this module is shown in Algorithm \ref{alg:rec}.

\begin{algorithm}[t]
\caption{Scholar Recommendation Module}
\label{alg:rec}
    \begin{algorithmic}[1]  
    \STATE \textbf{Input:} Graph $G = (V, E)$, target users $U$, max depth $T$
    \STATE \textbf{Output:} Recommendation list $S$
    \STATE Construct attribute subgraphs $\mathcal{G}$
    \STATE Generate augmented graph $G^{aug} \leftarrow$ Augment$(G, \mathcal{G})$
    \STATE Build behavioral path trees $\{\mathcal{T}^{bp}_{v_i}\}_{v_i \in V}$
    \FOR{$l = 1$ to $T$}
        \FOR{each $v_i \in V$}
            \STATE Extract paths $P_{v_i}^{(l-1)}$ from $\mathcal{T}_{v_i}^{bp}$
            \STATE Encode paths: $h_p^{(l-1)} \leftarrow$ PathEncoder$(P_{v_i}^{(l-1)})$
            \STATE Aggregate: $h_{v_i}^{(l)} \leftarrow$ HeteroAggregate$(h_{v_i}^{(l-1)}, h_p^{(l-1)})$
        \ENDFOR
    \ENDFOR
    \STATE Predict: $S \leftarrow$ PredictionHead$(h_{v_i}, U)$
    \RETURN $S$
    \end{algorithmic}
\end{algorithm}

\subsection{Path Mask Learning}

This module focuses on simplifying heterogeneous graph structures and identifying important paths for explanation. In large-scale academic graphs, nodes often possess numerous neighbors connected via diverse semantic relations. Searching across all multi-hop neighbors for potential explanatory paths is computationally expensive and prone to noise introduced by irrelevant or redundant edges. To mitigate these issues, we propose a two-stage strategy: (1) \textbf{Graph simplification via $k$-shell decomposition}, and (2) \textbf{Edge importance via mask learning}.

\subsubsection{\textbf{Graph simplification via $k$-shell decomposition.}}  
\begin{figure}[t]
    \centering
    \includegraphics[width=\linewidth]{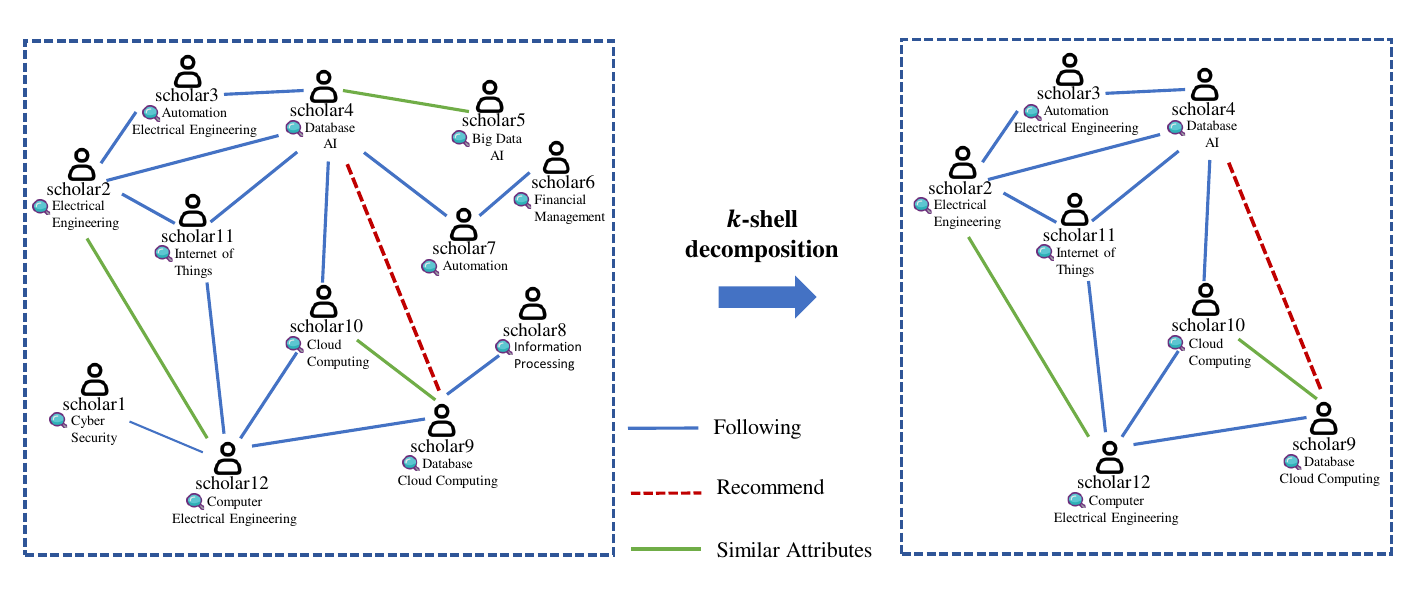}
    \caption{Example of 2-shell decomposition. Left: a 2-hop ego graph. Nodes with degree 1 (`Scholar 1', `Scholar 5', `Scholar 6', `Scholar 8') are removed first. Then `Scholar 7' is pruned recursively as its degree drops to 1. Right: the resulting 2-shell where all nodes have degree $\geq 2$.}
    \label{fig:k-shell-decomposition}
\end{figure}

For each scholar pair $(v_i, v_j)$, we first extract their $L$-hop ego graph $G^e = (V^e, E^e)$ from $G^{aug}$. It describes the local neighborhood structure used during message passing and provides the local context for generating potential explanations. We then apply $k$-shell decomposition~\citep{graph1984,k-shell2007} to reduce the size of $G^e$ by iteratively removing low-degree nodes. Specifically, the $k$-shell subgraph $G^k = (V^k, E^k)$ is the maximal connected subgraph in which each node has at least $k$ edges. The pruning process removes any node $v$ with degree $D_v < k$, and is applied recursively until all remaining nodes satisfy $D_v \geq k$. 
An example of 2-shell decomposition module is illustrated in Figure~\ref{fig:k-shell-decomposition}. In the left subfigure, nodes with degree 1 (`Scholar 1', `Scholar 5', `Scholar 6', `Scholar 8') are removed in the first iteration. As a result, the degree of `Scholar 7' drops to 1 and is subsequently pruned in the next iteration. The right subfigure shows the remaining core structure—i.e., the 2-shell—where all nodes have a degree of at least 2. This decomposition helps identify the densely connected core region around the target scholar, which serves as a foundation for path-based reasoning and improves the quality of selected paths.

In the context of Explainable AI, explanation fidelity requires that the generated paths faithfully reflect the underlying topological links relied upon by the recommendation model. Blindly dropping edges to reduce complexity often severs these critical links, damaging fidelity. We now provide a formal justification that our $k$-shell decomposition avoids this trap entirely, while reducing the computational search space.

Lemma 1 (Zero-Fidelity-Loss Pruning). Let $\Pi(v_s, v_t)$ denote the set of all valid multi-hop reasoning paths between a source scholar $v_s$ and a target scholar $v_t$ in graph $G$. Let $G^k$ denote the $k$-shell of $G$ with $k = 2$. Then $\Pi(v_s, v_t)$ is fully preserved in $G^k$, i.e., no valid reasoning path is eliminated by the decomposition.

Proof. Consider any path $p \in \Pi(v_s, v_t)$. Each intermediate node $v_i$ on $p$ must connect to at least its predecessor $v_{i-1}$ and its successor $v_{i+1}$. Since $v_i$ participates in at least one valid path, its degree in the full graph $G$ satisfies $\text{deg}_G(v_i) \geq 2$. The $2$-shell algorithm iteratively removes only nodes with degree less than 2. Since no intermediate node on any valid path has degree less than 2, no such node is removed, and therefore no valid path is disrupted. The source $v_s$ and target $v_t$ are additionally preserved by the algorithm's always-preserve constraint. 

Lemma 1 establishes that $k = 2$ decomposition is not a heuristic approximation but an exact and lossless operation with respect to the explanation path space. 
Beyond preserving all valid paths, $k$-shell decomposition also guarantees a substantial reduction in graph complexity. Classical results in random graph theory \citep{k-core1996} establish that the $k$-shell of a large graph with $n$ nodes and average degree $c$ retains only a constant fraction $\delta_V(c, k) < 1$ of nodes and $\delta_E(c, k) < 1$ of edges asymptotically. For an $L$-hop path search, this translates to an exponential reduction in the combinatorial search volume. While real-world scholar networks are not purely random graphs, this theoretical expectation is confirmed by our empirical measurements in Section \ref{sec:ablation_k-shell}, where we observe substantial node and edge reductions across all datasets.

\subsubsection{\textbf{Edge importance via mask learning.}}  
On the pruned graph, we learn a differentiable edge mask to identify important paths. Let $E_{r}^{k} = \{ e \in E^{k} \mid \omega(e) = r \}$ denote the set of edges of type $r \in R$, and $M = \{ M_r \}_{r=1}^{|R|}$ be the collection of learnable mask vectors, where $\omega$ is an edge type mapping function, $M_r \in \mathbb{R}^{|E_{r}^{k}|}$. The masked edge set is defined as:
\begin{equation}
    E^{k} \odot \sigma(M) = \bigcup_{r \in R} \left\{E_{r}^{k} \odot \sigma(M_r) \right\},
\end{equation}
where $\sigma(\cdot)$ is the sigmoid function and $\odot$ denotes element-wise multiplication. This yields a masked graph $G_m^{k} = (V^k, E^{k} \odot \sigma(M))$ where edges are softly selected based on learned importance weights.

Following the intuition that important paths should significantly influence recommendation outcomes, we optimize the edge mask by minimizing the recommendation performance degradation on the masked graph. Formally, we define a mask loss:
\begin{equation}
    \mathcal{L}_m(M) = - \log D_{\Psi}\left( Y = 1 \mid G_m^{k}, (v_i, v_j) \right),
    \label{eq: Loss-Rec}
\end{equation}
where $\Psi$ is the pretrained scholar recommendation model, and $D_{\Psi}(\cdot)$ denotes its predictive distribution over the recommendation label $Y$ given a masked graph and scholar pair. Maximizing this conditional likelihood encourages the mask to retain edges critical for accurate prediction.

\begin{algorithm}[t]
\caption{Recommendation Explanation Module}
\label{alg:explan}
\begin{algorithmic}[1]  
\STATE \textbf{Input:} Augmented graph $G^{aug}$, target scholars $U$, recommendation results $S$, path length $T$, scholar model $\Psi(\cdot, \cdot)$, max iterations $I$
\STATE \textbf{Output:} Explanatory path set $\mathbb{P}$
\STATE Extract $L$-hop ego graph $G^e$ for $(v_i, v_j)$
\STATE $G^k \leftarrow$ Apply $k$-shell decomposition on $G^e$
\STATE Initialize mask $M(0)$
\STATE $i \leftarrow 0$
\WHILE{$M(i)$ not converged \AND $i < I$}
    \STATE Compute $\mathcal{L}_m(M)$ using Equation~\ref{eq: Loss-Rec}
    \STATE Compute edge importance scores $\mathcal{E}_p$ (see Eq.~\ref{eq: Score})
    \STATE Extract candidate paths $\mathbb{P}_c$ based on shortest paths with highest $\mathcal{E}_p$
    \STATE Compute optimization loss $\mathcal{L}_{opt}(M)$ using Equation~\ref{eq: Loss-opt}
    \STATE Update $M(i+1)$ based on $\mathcal{L}_m(M)$ and $\mathcal{L}_{opt}(M)$
    \STATE $i \leftarrow i + 1$
\ENDWHILE
\STATE $\mathbb{P} \leftarrow$ Run shortest-path search over top-scored edges
\RETURN $\mathbb{P}$
\end{algorithmic}
\end{algorithm}

\subsection{Explanation Generation}

While the loss $\mathcal{L}_{rec}$ effectively captures the structural factors driving predictions, directly using these critical edges often leads to redundant or generic paths, especially in dense networks with high-degree scholars (i.e., hubs). 
To enhance explainability, this module filters and refines candidate paths by incorporating path length and node centrality, ultimately generating compact and informative explanation paths.
\subsubsection{\textbf{Path selection via explainability scoring.}}
Although mask learning effectively identifies important edges, directly combining them into connective paths can yield overly lengthy or insufficiently specific explanations. This issue is particularly prominent in large ASNs, where prolific scholars usually have rich attribute profiles. They engage in diverse collaborations and exhibit broad research interests. Consequently, the generated paths often produce highly redundant explanations between users and scholars, which obscures key insights and complicates explainability.
To prioritize high-quality explanation paths, we define a scoring function that considers both edge importance and node centrality:
\begin{equation}
    \mathcal{E}_{p} = \sum_{e = (v_i, v_j) \in p} \left( \log \sigma(M_{\omega(e)}^e) - \log (\text{Centrality}(v_j)) \right),
    \label{eq: Score}
\end{equation}
where $\text{Centrality}\left(\cdot \right)$ refers to the degree centrality of node. This function favors paths that consist of important edges connected to less generic scholars, thereby enhancing the specificity and clarity of explanations. We then extract a candidate path set $\mathbb{P}_c$ by selecting the shortest paths with the highest $\mathcal{E}_p$ scores.

\subsubsection{\textbf{Path refinement via explanation-oriented optimization.}}
To further enhance explanation quality, we refine the mask weights by optimizing a loss that emphasizes the inclusion of high-scoring paths and suppresses uninformative ones:
\begin{equation}
    \mathcal{L}_{opt}(M) = - \sum_{\omega(e) \in R} \left( \alpha \sum_{e \in \mathbb{P}_c} M_{\omega(e)}^e - \beta \sum_{e \notin \mathbb{P}_c} M_{\omega(e)}^e \right),
    \label{eq: Loss-opt}
\end{equation}
where $\alpha$ and $\beta$ are trade-off hyperparameters. This optimization encourages the mask to highlight edges belonging to interpretable paths while down-weighting irrelevant ones. After convergence, we apply a shortest-path algorithm over the optimized masked graph to extract the final explanation path set $\mathbb{P}$, constrained by path length and relevance. The pseudocode of this module is shown in Algorithm \ref{alg:explan}.

A natural concern arising from the attribute-sharing graph augmentation is the potential explosion in the number of connective paths. Our framework addresses this through a multi-level automated pruning mechanism that requires no manual filtering. At the topology level, $k$-shell decomposition deterministically removes nodes that cannot serve as intermediate bridging nodes on any shortest path, providing a rigorous topological guarantee. At the semantic level, the edge mask learning loss $\mathcal{L}_m$ automatically suppresses the weights of edges that do not contribute meaningful collaborative filtering signals to the recommendation decision. At the optimization level, the combined score function $\mathcal{E}_{p}$ and optimization loss $\mathcal{L}_{opt}$ enforce extreme sparsity on the mask distribution, retaining only a small set of high-quality explanation paths. Together, these three levels ensure that the output explanation set is both concise and reasonable without any human intervention.

\subsection{Complexity Analysis}
\label{sec:complexity}
For the recommendation section, the time complexity for constructing all behavioral path trees is $O(\left| V \right| \cdot \bigtriangleup ^{L})$, where $L$ is the maximum depth, $\bigtriangleup$ represents the average node degree. Since each node has $O( \bigtriangleup ^{L} )$ paths in the worst case, the path encoding complexity per layer is $O( \left| V \right| \cdot \bigtriangleup ^{L} \cdot L)$. The path aggregation operation combines information from multiple relation types, its complexity is $O( \left| V \right| \cdot \bigtriangleup ^{L} \cdot \left| R \right|)$. Thus, the time complexity of scholar recommendation is $O( L \cdot \left| V \right| \cdot \bigtriangleup ^{L} \cdot (L + \left| R \right|))$. While this appears exponential in $L$, we note that in real-world academic graphs, $\bigtriangleup$ is typically small due to the sparse connectivity and the simple path constraint. Furthermore, we employ sampling strategies during training to maintain computational feasibility. The memory complexity is primarily determined by storing path representations and is $O( \left| V \right| \cdot \bigtriangleup ^{L})$.
For the explanation section, the $k$-shell decomposition step is linear in $\left|E^e\right|$. The mask learning for path generation has complexity $\left|G^k\right| \cdot T$, where $T$ is the mask learning epochs. So the time complexity of explanation generation is $O(\left|E^e\right| + \left|G^k\right| \cdot T)$.

\section{Experiments}
\label{sec:exp}
In this section, we conduct experiments on \textit{Recommendation Accuracy} and \textit{Model Explainablity} to evaluate the effectiveness of the proposed PAGSL model.

\subsection{Experimental Settings}
\subsubsection{\textbf{Datasets}}
\begin{table}[htp]
    \centering
    \begin{threeparttable} 
    \caption{Basic information of the experimental datasets.}
        \begin{tabular}{ccccc}
            \toprule 
            Datasets & \#Nodes & \#Train edges & \#Test edges & \#Sparsity \\
            \midrule 
            Scholat & 9,537 & 128,924 & 32,231 & 99.65\% \\ 
            AMiner & 36,770 & 110,880 & 35,952 & 99.98\% \\
            ResearchGate & 2,659 & 21,848 & 5,462 & 99.23\% \\
            \bottomrule 
        \end{tabular}
        \label{Dataset Information}
    \end{threeparttable}
\end{table}

We conduct experiments on three real-world academic datasets: Scholat\footnote{\url{https://www.scholat.com/}}, AMiner\footnote{\url{http://www.aminer.cn/aminernetwork}}, and ResearchGate\footnote{\url{https://www.researchgate.net/}}. The basic information is summarized in Table \ref{Dataset Information}.
\begin{itemize}
\item \textbf{Scholat}: The Scholat dataset originates from an emerging vertical academic social networking platform specifically designed for researchers in China. It contains user-follow relationships and academic profiles of 9,537 Chinese researchers. We standardize heterogeneous textual attributes using machine translation and segmentation, remove duplicates, and construct unified attribute descriptions.

\item \textbf{AMiner}: The Aminer dataset comprises multiple relational networks, such as citation networks, academic social networks, or topic paper author networks. We select papers from 2008 to 2014, excluding those with fewer than six authors or ten references, and filter out authors lacking affiliation data or having only one publication. The final dataset contains 9,510 papers and 36,770 authors, with a temporal split at 2012 for training and testing.

\item \textbf{ResearchGate}: The ResearchGate dataset is an academic social networking platform designed to help researchers share, discuss and find academic resources. It is preprocessed following~\citep{GESRec2021}, resulting in 2,659 researchers, 13,241 publications, and 27,310 follow relations.

\end{itemize}

\begin{figure*}
    \begin{subfigure}{0.32\textwidth}
        \centering
        \includegraphics[width=\linewidth]{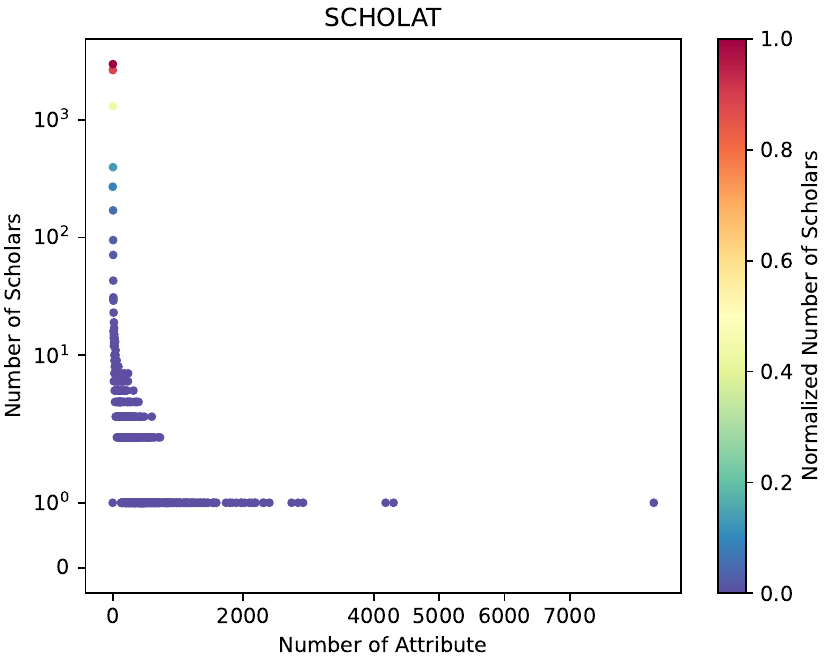}
        \caption{Scholat}
        \label{fig:scholat-attribute}
    \end{subfigure}
    \hfill
    \begin{subfigure}{0.32\textwidth}
        \centering
        \includegraphics[width=\linewidth]{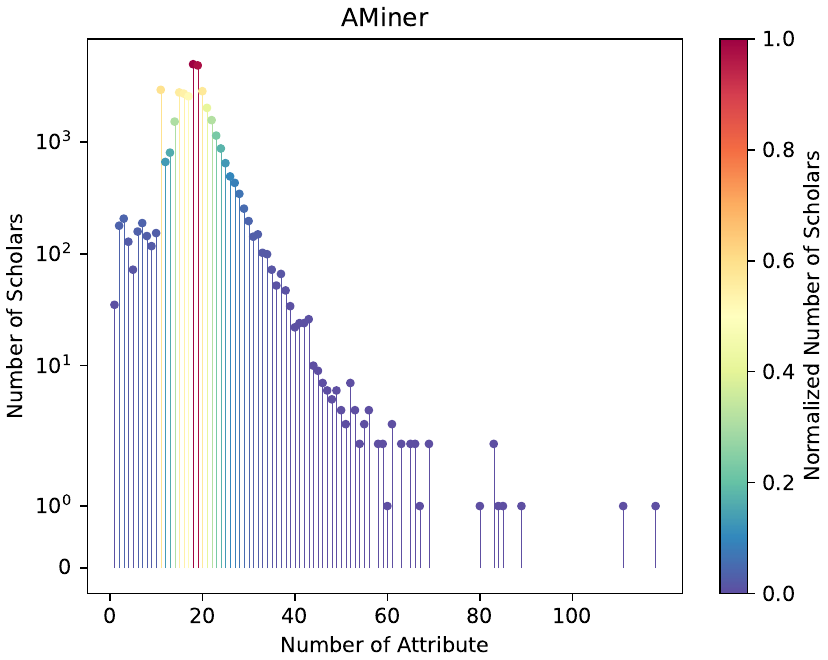}
        \caption{AMiner}
        \label{fig:aminer-attribute}
    \end{subfigure}
    \hfill
    \begin{subfigure}{0.32\textwidth}
        \centering
        \includegraphics[width=\linewidth]{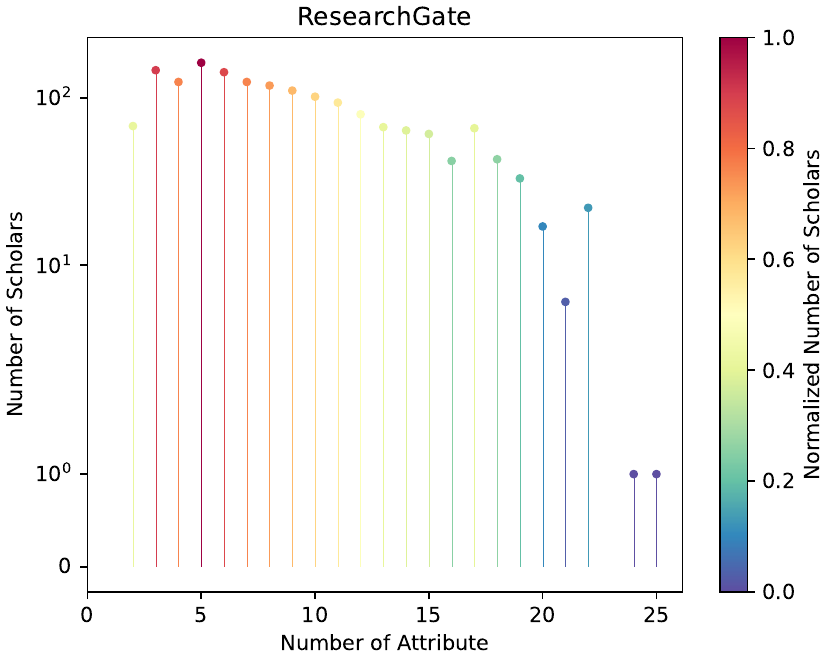}
        \caption{ResearchGate}
        \label{fig:researchgate-attribute}
    \end{subfigure}

    \caption{The distribution of scholars attributes across three datasets. The plots illustrate a long-tail distribution, particularly in the Scholat and AMiner datasets. The vast majority of scholars possess a highly limited number of attributes, concentrated in the lower ranges, while only a few hub scholars exhibit rich attribute sets.}
    \label{Fig: Attribute Statistics}
\end{figure*}

The attribute characteristics of different scholars differ greatly, as shown in Figure \ref{Fig: Attribute Statistics}, which shows the relationship between the number of scholars and the attribute counts on three datasets. Compared to the ResearchGate, the attribute distribution of scholars on Scholat and AMiner datasets is more imbalanced, showing a clear long tail distribution. It can be noted that the vast majority of scholars have a limited number of attributes and are concentrated in lower attribute ranges, while as the number of attributes increases, the number of scholars rapidly decreases, and only a few scholars have more attributes. To balance the attribute differences among scholars and utilize this attribute information for scholar recommendation, this research preserves the shared attribute relationships in the form of a augmented graph, and the implementation steps are described in Section \ref{sec:G_aug}.

\subsubsection{\textbf{Baselines and Evaluation Metrics}}
For \textit{Recommendation Accuracy}, we consider methods across four categories. NeuMF~\citep{NeuMF2017}, NGCF~\citep{NGCF2019}, and LightGCN~\citep{LightGCN2020} are collaborative filtering-based models. GNN-based recommenders include GAT~\citep{GAT2018}, GraphSAGE~\citep{GraphSAGE2017}, and SIGMA~\citep{SIGMA2025}. Contrastive learning methods such as SGL~\citep{SGL2021}, HGCL~\citep{HGCL2023} and Pone-GNN~\citep{Pone-GNN2025}. Heterogeneous graph learning models include PathNN~\citep{PathNN2023}, DiffKG~\citep{DiffKG2024} and GC-HGNN~\citep{GC-HGNN2025}, LLM-enhanced models include LLM4REC~\citep{LLM4REC2025}. To assess \textit{Model Explainability}, we include two general-purpose GNN explanation methods adapted to recommendation: GNNExplainer~\citep{GNNExplainer2019}, and PGExplainer~\citep{PGExplainer2020}.

We adopt classification metrics (Precision, Recall, F1-score, Accuracy and AUC) and ranking metrics (Hit Ratio (HR@$K$) and Normalized Discounted Cumulative Gain (NDCG@$K$) at $K \in \{5, 10\}$) to evaluate recommendation performance. For explainability evaluation, we follow~\citet{PGExplainer2020} and compute AUC by comparing learned edge masks against ground-truth edges. Edges in the annotated ground truth are treated as positives, while others are negatives; predicted mask weights are used as scores.

\begin{table}[!htpb]
\centering
\caption{Structured novelty comparison between PAGSL and recent SOTA methods.}
\label{tab:novelty_comparison}
\renewcommand{\arraystretch}{1.3}
\scriptsize 
\begin{tabular}{@{} c *{7}{c} @{}}
\toprule
\textbf{Dimension} & HGCL & Pone-GNN & DiffKG & GC-HGNN & CaGE & TMER-RL & \textbf{PAGSL} \\
\midrule
\multicolumn{8}{@{}l}{\textit{--- Core Technical Capabilities ---}} \\
\quad Encoding Mechanism & \multicolumn{2}{c}{\makecell{Multi-View \\ Contrastive Encoding}} & \multicolumn{2}{c}{\makecell{Generative- \\ Augmented Encoding}} & \makecell{Causal \\ Masked Path} & \makecell{Attentive \\ Path} & \makecell{\textbf{Path-Tree} \\ \textbf{Branching}} \\
\quad Topology Refinement & $\times$ & $\times$ & \checkmark & \checkmark & \checkmark & $\times$ & \textbf{\checkmark} \\
\quad Learnable Edge Mask & $\times$ & $\times$ & $\times$ & \checkmark & \checkmark & $\times$ & \textbf{\checkmark} \\
\quad Predefined Meta-path & $\times$ & $\times$ & $\times$ & \checkmark & \checkmark & \checkmark & \textbf{$\times$} \\
\midrule
\multicolumn{8}{@{}l}{\textit{--- Explainability \& Robustness ---}} \\
\quad Explanation Type & \multicolumn{4}{c}{$-$} & \makecell{Meta-path \\ Explanations} & \makecell{Sequential \\ Paths} & \makecell{\textbf{Adaptive} \\ \textbf{Paths}} \\
\quad Augmentation Serves Explanation & $\times$ & $\times$ & $\times$ & $\times$ & $\times$ & $\times$ & \textbf{\checkmark} \\
\quad Resistance to Hub-Node Bias & $\times$ & $\times$ & $\times$ & $\times$ & $\times$ & $\times$ & \textbf{\checkmark} \\
\bottomrule
\end{tabular}
\end{table}

Table \ref{tab:novelty_comparison} provides a structured comparison of our framework with recent representative methods across key technical dimensions. While existing methods typically address either recommendation quality or explanation generation, our framework jointly addresses both within a unified pipeline. Specifically, PAGSL introduces three key structural advantages. First, our attribute-sharing graph augmentation is unique in serving a dual purpose: improving recommendation accuracy and creating semantically typed edges for explanation paths. Second, it employs a path-tree encoding that naturally preserves the branching propagation of academic influence. Finally, to combat the "hub-node bias" prevalent in scale-free networks, we propose an automated multi-level pruning pipeline. By coupling $k$-shell decomposition with a centrality-penalized mask learning objective, PAGSL refines noisy topologies and actively suppresses uninformative hub nodes without requiring heuristic manual filtering.

\subsubsection{\textbf{Implementation Details}}
We adopt an edge-level split strategy to ensure rigorous evaluation of generalization performance. All performance metrics are evaluated on completely unseen testing sets that were strictly isolated from the model during the training and validation phases. 
For the Scholat and ResearchGate datasets, we randomly partitioned the interaction pairs into training, validation, and testing sets with a strict ratio of 6:2:2. 
For the AMiner dataset, we adopt a chronological splitting strategy. Specifically, interactions between 2008 and 2012 are utilized to construct the training sets, interactions from 2013 to 2014 are utilized to construct the unseen testing set.

Embedding dimensions are set to 128 for Scholat and ResearchGate, and 256 for AMiner. The recommendation module is trained for 200 epochs, and the explanation module for 20. 
Negative samples are selected from $\{1, 3, 5\}$ via grid search.
The recommendation module uses Adam with learning rates of 0.07 (Scholat), 0.002 (AMiner), and 0.001 (ResearchGate). Prediction heads are chosen from {\texttt{dot}, \texttt{mlp}, \texttt{cos}} based on validation. In the explanation module, the learning rate is set to 0.1. We use 2-hop ego graphs and extract up to 40 paths of maximum length 5. The $k$-shell number is fixed at 2. Regularization weights $\alpha$ and $\beta$ are selected from \{$0.1, 0.5, 1.0, 1.5, 2.0$\}. 

To ensure a rigorous and fair comparison, all baseline models are implemented using the optimal configurations reported in the original literature. For example, for the Pone-GNN model, the architecture was configured with four GNN layers and a filtering threshold of $0.5$, with the initial learning rate established at $5 \times 10^{-3}$. For LLM4REC, we configured the token embedding dimension to 768, the maximum sequence length to 1,024, and the natural language vocabulary size to 50,257. Other competitive benchmarks were similarly parameterized according to their official implementations or standard protocols, ensuring that all models are evaluated under their most favorable conditions.
There is no input information mismatch between our PAGSL framework and the baseline models. All models are provided with identical access to both structural interaction links and scholar attribute information.

\begin{table*}[h]
    \centering
    \begin{threeparttable}
    \caption{Experimental results on three datasets with classification metrics.}
    \label{tab:scholat_transposed}
    
    \renewcommand{\arraystretch}{1.3}
    
    \scriptsize
    \setlength{\tabcolsep}{3.0pt}
        \begin{tabular*}{\textwidth}{@{\extracolsep{\fill}} l ccccc ccccc ccccc}
            \toprule
            \multirow{2}{*}{\textbf{Method}} & \multicolumn{5}{c}{\textbf{Scholat}} & \multicolumn{5}{c}{\textbf{AMiner}} & \multicolumn{5}{c}{\textbf{ResearchGate}} \\
            \cmidrule(lr){2-6} \cmidrule(lr){7-11} \cmidrule(lr){12-16}
            & Pre. & Rec. & F1 & ACC & AUC & Pre. & Rec. & F1 & ACC & AUC & Pre. & Rec. & F1 & ACC & AUC \\
            \midrule
            NeuMF & 0.7683 & 0.7726 & 0.7705 & 0.7699 & 0.8610 & 0.5482 & 0.7193 & 0.6222 & 0.5670 & 0.6169 & 0.7876 & 0.4848 & 0.6002 & 0.4944 & 0.5166 \\
            NGCF & 0.6453 & 0.9542 & 0.7700 & 0.7149 & 0.7693 & 0.5253 & 0.8365 & 0.6453 & 0.5402 & 0.5403 & 0.5637 & 0.8460 & 0.6766 & 0.5957 & 0.5959 \\
            LightGCN & 0.9159 & 0.6515 & 0.7614 & 0.7958 & 0.8073 & 0.9223 & 0.1627 & 0.2765 & 0.5745 & 0.5749 & 0.7364 & 0.3089 & 0.4352 & 0.5991 & 0.6047 \\
            GAT & 0.8006 & 0.6964 & 0.7449 & 0.7615 & 0.8106 & 0.8698 & 0.7332 & 0.7956 & 0.8117 & 0.7667 & \underline{0.8282} & 0.7432 & 0.7834 & 0.7284 & 0.7657 \\
            GraphSAGE & 0.6521 & 0.8291 & 0.7300 & 0.6933 & 0.7687 & 0.8645 & 0.8270 & 0.8454 & 0.8487 & 0.9183 & 0.6570 & 0.8555 & 0.7460 & 0.7062 & 0.7763 \\
            SIGMA & \underline{0.9531} & 0.9372 & 0.9451 & 0.9455 & 0.9781 & 0.9528 & 0.8459 & 0.8962 & 0.9020 & 0.9314 & 0.8093 & 0.7862 & 0.7976 & 0.8005 & 0.8629 \\
            SGL & 0.6758 & 0.9600 & 0.7932 & 0.7498 & 0.9018 & 0.7038 & 0.9482 & 0.8002 & 0.7514 & 0.7115 & 0.6354 & 0.7121 & 0.6716 & 0.6517 & 0.7032 \\
            HGCL & 0.6799 & \textbf{0.9738} & 0.8003 & 0.7521 & 0.9502 & 0.8355 & 0.8055 & 0.8202 & 0.8474 & 0.9524 & 0.5546 & 0.8790 & 0.6801 & 0.5866 & 0.6199 \\
            Pone-GNN & 0.9313 & 0.9570 & 0.9440 & 0.9432 & 0.9709 & 0.8239 & 0.5811 & 0.6815 & 0.7285 & 0.7109 & 0.6025 & 0.7519 & 0.6689 & 0.6279 & 0.7016 \\
            PathNN & 0.8285 & 0.8981 & 0.8619 & 0.8561 & 0.9382 & 0.8856 & 0.8663 & 0.8758 & 0.8772 & 0.9451 & 0.7410 & 0.7469 & 0.7440 & 0.7429 & 0.8151 \\
            DiffKG & 0.9193 & 0.9621 & 0.9402 & 0.9388 & 0.9793 & 0.7633 & 0.8462 & 0.8026 & 0.7918 & 0.8667 & 0.7752 & \textbf{0.9260} & 0.8439 & 0.8287 & 0.8678 \\
            GC-HGNN & 0.8785 & 0.9513 & 0.9135 & 0.9099 & 0.9518 & 0.8115 & 0.8056 & 0.8085 & 0.8092 & 0.8752 & 0.5944 & \underline{0.9017} & 0.7166 & 0.6432 & 0.7074 \\
            LLM4REC & 0.8664 & 0.9440 & 0.9035 & 0.8992 & 0.9541 & 0.9209 & 0.9122 & 0.9165 & 0.9169 & 0.9716 & 0.8036 & 0.7835 & 0.7934 & 0.7960 & 0.8684 \\
            
            \midrule
            PAGSL(\texttt{dot}) & 0.9323 & 0.9060 & 0.9190 & 0.9201 & 0.9655 & \underline{0.9537} & 0.9278 & 0.9405 & 0.9414 & 0.9827 & 0.8102 & 0.8734 & 0.8406 & 0.8344 & 0.9054 \\
            PAGSL(\texttt{mlp}) & 0.9367 & 0.9584 & \underline{0.9474} & \underline{0.9468} & \underline{0.9818} & 0.9367 & \underline{0.9596} & \underline{0.9480} & \underline{0.9474} & \textbf{0.9893} & 0.8152 & 0.8918 & \textbf{0.8518} & \underline{0.8448} & \underline{0.9108} \\
            PAGSL(\texttt{cos}) & \textbf{0.9744} & \underline{0.9696} & \textbf{0.9720} & \textbf{0.9721} & \textbf{0.9899} & \textbf{0.9874} & \textbf{0.9694} & \textbf{0.9783} & \textbf{0.9785} & \underline{0.9856} & \textbf{0.8520} & 0.8457 & \underline{0.8488} & \textbf{0.8494} & \textbf{0.9233} \\
            \bottomrule
        \end{tabular*}
  \end{threeparttable}
\end{table*}

\begin{table*}[!ht]
    \centering
    \begin{threeparttable}
    \caption{Experimental results on three datasets with ranking metrics.}
    \label{tab:recommendation with HR and NDCG}
    
    \renewcommand{\arraystretch}{1.3}
    
    \scriptsize
    \setlength{\tabcolsep}{2.0pt}
    
        \begin{tabular*}{\textwidth}{@{\extracolsep{\fill}} >{}l *{12}{>{}c} } 
            \toprule
            \multirow{2}{*}{\textbf{Method}} & \multicolumn{4}{c}{\textbf{Scholat}} & \multicolumn{4}{c}{\textbf{AMiner}} & \multicolumn{4}{c}{\textbf{ResearchGate}} \\
            \cmidrule(lr){2-5} \cmidrule(lr){6-9} \cmidrule(lr){10-13}
            & HR@5 & NDCG@5 & HR@10 & NDCG@10 & HR@5 & NDCG@5 & HR@10 & NDCG@10 & HR@5 & NDCG@5 & HR@10 & NDCG@10 \\
            \midrule
            NeuMF & 0.1557 & 0.0554 & 0.1987 & 0.0580 & 0.1372 & 0.0536 & 0.1819 & 0.0619 & 0.1295 & 0.0396 & 0.1891 & 0.0439  \\
            NGCF & 0.1621 & 0.0916 & 0.1656 & 0.0893 & 0.1339 & 0.0661 & 0.1543 & 0.0720 & 0.1244 & 0.0560 & 0.1295 & 0.0472 \\
            LightGCN & 0.1812 & 0.0764 & 0.3592 & 0.1196 & 0.1829 & 0.0946 & 0.2022 & 0.1009 & 0.1762 & 0.0759 & 0.2228 & 0.0644 \\
            GAT & 0.1903 & 0.0859 & 0.2769 & 0.1045 & 0.2193 & 0.1221 & 0.3065 & 0.1521 & 0.1736 & 0.0637 & 0.3290 & 0.0893 \\
            GraphSAGE & 0.1636 & 0.0689 & 0.2641 & 0.0884 & 0.2672 & 0.1150 & 0.3640 & 0.1387 & 0.1425 & 0.0497 & 0.3342 & 0.0744 \\
            SIGMA & 0.2118 & 0.1138 & 0.2547 & 0.1260 & 0.2637 & 0.1700 & 0.3129 & 0.1848 & 0.2043 & 0.0704 & 0.2536 & 0.0862 \\
            SGL & 0.1520 & 0.0936 & 0.2737 & 0.1326 & 0.1449 & 0.1179 & 0.1576 & 0.1220 & 0.1023 & 0.0758 & 0.1270 & 0.0838 \\
            HGCL & 0.1743 & 0.1116 & 0.2814 & 0.1460 & 0.3087 & \textbf{0.1796} & 0.3769 & \textbf{0.2013} & 0.1190 & 0.0672 & 0.1701 & 0.0833 \\
            Pone-GNN & 0.2599 & \textbf{0.1335} & 0.3984 & \underline{0.1565} & 0.1713 & 0.1585 & 0.1871 & 0.1635 & 0.2098 & 0.0797 & 0.2483 & 0.0920 \\
            PathNN & 0.1780 & 0.0608 & 0.2979 & 0.0825 & 0.1921 & 0.1375 & 0.2859 & 0.1605 & 0.1451 & 0.0435 & 0.2668 & 0.0557 \\
            DiffKG & 0.2485 & 0.1166 & 0.3683 & 0.1552 & 0.1669 & 0.1215 & 0.2259 & 0.1405 & 0.1911 & 0.0668 & 0.2437 & 0.0805 \\
            GC-HGNN & 0.2133 & \underline{0.1290} & 0.2955 & 0.1556 & 0.2137 & 0.1481 & 0.3231 & 0.1831 & 0.1098 & 0.0689 &
            0.1781 & 0.0910 \\
            LLM4REC & 0.2352 & 0.1119 & 0.3158 & 0.1504 & 0.3063 & 0.1601 & 0.4086 & 0.1816 & 0.2150 & 0.0652 & 0.3008 & 0.0897 \\
            
            \midrule
            PAGSL(\texttt{dot}) & \textbf{0.2839} & 0.1200 & \underline{0.4397} & 0.1540 & \underline{0.3413} & 0.1556 & \underline{0.4490} & 0.1844 & \underline{0.2409} & \underline{0.0848} & \underline{0.4430} & \textbf{0.1121} \\
            PAGSL(\texttt{mlp}) & \underline{0.2759} & 0.1175 & 0.4020 & 0.1466 & 0.3270 & 0.1522 & 0.4533 & 0.1861 & \textbf{0.2539} & \textbf{0.0872} & 0.3860 & 0.0937 \\
            PAGSL(\texttt{cos}) & 0.2711 & 0.1221 & \textbf{0.4457} & \textbf{0.1647} & \textbf{0.3567} & \underline{0.1703} & \textbf{0.4495} & \underline{0.1935} & 0.2358 & 0.0818 & \textbf{0.4637} & \underline{0.1095} \\
            \bottomrule
        \end{tabular*}
  \end{threeparttable}
\end{table*}

\subsection{Experimental Results}
\subsubsection{\textbf{Recommendation Performance}}
Table~\ref{tab:scholat_transposed} and Table~\ref{tab:recommendation with HR and NDCG} present the performance comparison between the proposed PAGSL model and baselines across three benchmark datasets. We report results for three variants of PAGSL that differ in their choice of prediction head: PAGSL(\texttt{dot}), PAGSL(\texttt{cos}), and PAGSL(\texttt{mlp}). The major findings can be summarized as follows.
\begin{itemize}[noitemsep, topsep=2pt, leftmargin=*]

    \item The overall results consistently demonstrate the robustness of PAGSL across varying levels of data sparsity. On the Scholat dataset, PAGSL variants achieve the best performance across most metrics, with PAGSL(\texttt{cos}) securing the highest HR@10 and NDCG@10. While specific contrastive learning baselines like Pone-GNN exhibit highly competitive ranking precision, PAGSL maintains a clear advantage in hit ratios and overall classification tasks. This highlights the advantage of our path-aware relational modeling in capturing nuanced collaboration semantics in academic social networks.

    \item On the highly sparse AMiner dataset, PAGSL exhibits strong robustness, outperforming contrastive learning and GNN-based baselines in F1-score, AUC, and Hit Ratios. Although the HGCL baseline achieves the highest NDCG scores on this specific dataset, Pone-GNN exhibits a substantial decline in efficacy, underscoring the limitations of data augmentation-based contrastive learning methods in sparse environments. Furthermore, while LLM4REC demonstrates competitive performance by leveraging latent semantics, it is still surpassed by PAGSL in most metrics. This indicates that while LLMs provide strong semantic foundations, the combination of graph augmentation and task-aligned academic behavior path generation in PAGSL offers a more robust solution for alleviating the sparse user-item interactions challenges.

    \item  For the relatively dense ResearchGate dataset, PAGSL(\texttt{dot}) and PAGSL(\texttt{cos}) achieve the best scores across all HR and NDCG metrics, alongside top F1-score and AUC performance. This indicates the model's ability to exploit academic interactions while remaining resilient to noise in dense graph structures.
    Moreover, collaborative filtering methods exhibit considerable performance degradation due to their inability to incorporate structural information. Heterogeneous graph methods such as DiffKG and GC-HGNN are also less effective, potentially due to the inclusion of irrelevant relations from general knowledge graphs or predefined meta-paths, which introduce noise into the learning process.

    \item In contrast, the graph relationship learning module in PAGSL adaptively generates academic behavior paths, filters redundant or noisy relations, and aggregates informative path representations. 
    PAGSL effectively overcomes this limitation by jointly modeling structural semantics from knowledge graphs and implicit user behavior patterns, leading to superior recommendation performance.
\end{itemize}

\begin{table}[t]
    \centering
    \begin{threeparttable}
    \caption{Statistics on AUC scores of different explainers.}
    \label{table 3}
        \begin{tabular}{cccccccccc}
            \toprule 
            Datasets & Metrics & GNNExp-SR & PGExp-SR & PAGSL \\
            \midrule 
            Scholat & AUC & 0.6584 & 0.5010 & \textbf{0.8458} \\ 
             
            AMiner & AUC & 0.5503 & 0.5076 & \textbf{0.8571} \\
                    
            ResearchGate & AUC & 0.5847 & 0.4651 & \textbf{0.8010} \\
             
            \bottomrule 
        \end{tabular}
    \end{threeparttable}
\end{table}

\subsubsection{\textbf{Explanation Quality}}

To quantitatively assess explanation fidelity, we compare PAGSL with GNNExplainer and PGExplainer. 
Higher AUC indicates better alignment between identified explanatory edges and the model’s internal decision process\footnote{The AUC here is inconsistent with the previous one, which is a quantification of explainability.}.
As shown in Table~\ref{table 3}, PAGSL consistently outperforms the baselines on all datasets. This superiority derives from its path-based approach, which activates relational paths via graph augmentation and learns heterogeneous edge masks to identify critical behavior paths, resulting in more accurate and faithful explanations.

\begin{figure}[t]
	\centering
		\includegraphics[scale=.68]{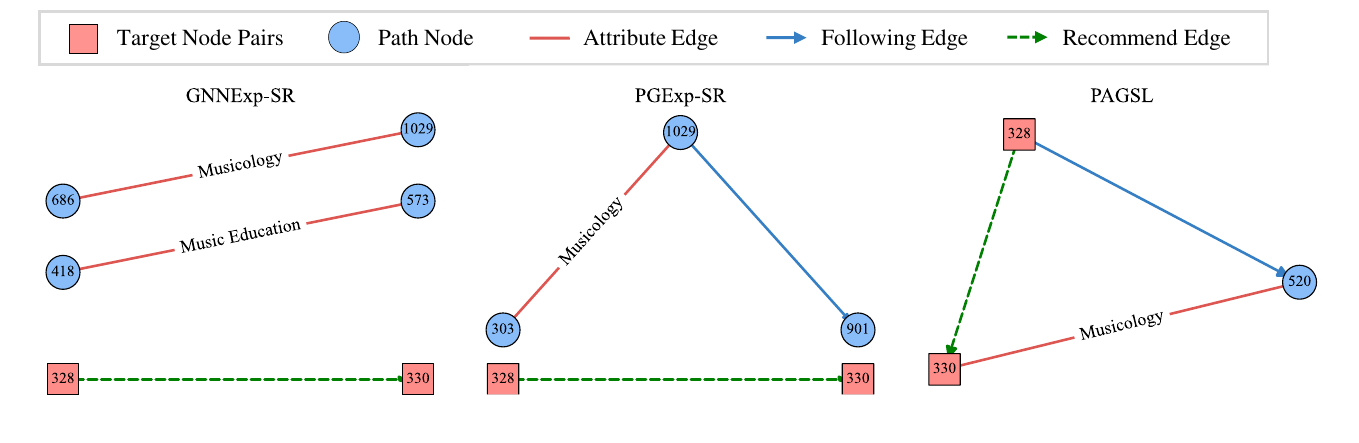}
	\caption{Comparison of explanations generated by different explainers for the scholar recommendation. The baseline methods produce disjointed subgraphs that fail to connect the target scholars. In contrast, our PAGSL framework generates a continuous and logical reasoning path. It clearly connects `Scholar 328' to the recommended follows `Scholar 330'via a shared interest in 'Musicology', providing a highly intuitive explanation.}
    \vspace{-0.3cm}
	\label{fig:Explain_example_1}
\end{figure}

Subsequently, we visualize the explanations derived from scholar recommendations on the Scholat dataset to facilitate comparative analysis. As illustrated in Figure \ref{fig:Explain_example_1}, the model recommends `Scholar 330' to `Scholar 328'. Both GNNExplainer and PGExplainer generate explanatory subgraphs for the node pair; however, the subgraphs produced exhibit disconnected edges from the target nodes (`Scholar 328', `Scholar 330'), thereby hindering explainability of the recommendation. In comparison, the PAGSL model provides an effective path explanation: `Scholar 328' follows `Scholar 520', reflecting an interest in topics associated with the latter. Given that both `Scholar 520' and `Scholar 330' share the research interest `Musicology', they may have more commonalities in academic research. Therefore, we will recommend `Scholar 330' to User `Scholar 328'. Under the budget condition of two connected edges, the explanation generated by our method has a better explainability and is easier for users to understand.

\begin{figure*}[t]
	\centering
		\includegraphics[scale=.65]{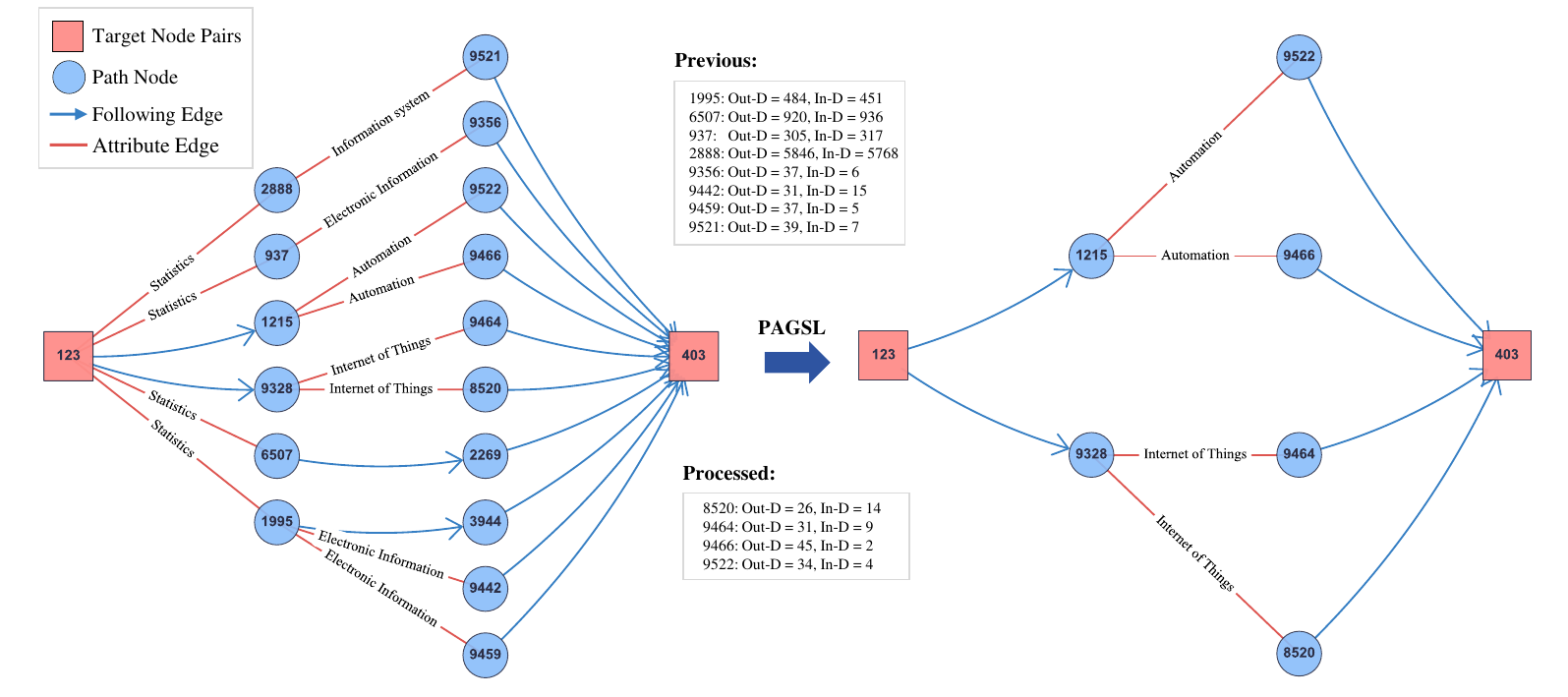}
	\caption{A case study of explanation paths selected by PAGSL. The left is the paths from mask learning, and the right is the optimized explanation paths via scoring function. The final paths are concise and informative, and effectively explains the recommendation results from the perspectives of attribute sharing and collaborative filtering.}
        \vspace{-0.3cm}
	\label{fig:Explain_example_2}
\end{figure*}

To further illustrate the capability of our approach in selecting concise and informative paths, an additional scholar recommendation example is provided in Figure \ref{fig:Explain_example_2}, where `Scholar 123' is the source node and `Scholar 403' is the recommended target. The top-4 path generated by PAGSL can be explained through attribute sharing and collaborative filtering. This path passes through two nodes (`Scholar 1216' and `Scholar 9328') that the source `Scholar 123' is interested in, as well as their corresponding attribute sharing relationships, and reaches four nodes (`Scholar 9466', `Scholar 9522', `Scholar 8520', `Scholar 9464') that may have an intimate relationship with `Scholar 123'. Then `Scholar 123' may also be interested in the scholars who are followed by these four nodes. Before path selection, there were many explainable paths between `Scholar 123' and `Scholar 403'. However, `Scholar 1995', `Scholar 6507', `Scholar 937', and `Scholar 2888' are high-degree universal connectors, and and paths passing through them are associated with generic edges such as `Statistics' and `Electronic Information', resulting in lower informative explanations. Therefore, to obtain a more concise and informative explanation path, this method will remove such high-degree nodes and their paths, so that the final explanation path passes through fields such as `Internet of Things' and `Automation', rather than general fields such as `Statistics' and `Electronic Information'.

\subsection{Computational Efficiency}
While Section \ref{sec:complexity} provides the theoretical time complexities of our framework, it is essential to evaluate the concrete empirical runtime to demonstrate real-world scalability. To provide a comprehensive evaluation, we decoupled the runtime into the Rec. Phase (seconds per epoch for the recommendation task) and the Explain. Phase (seconds per query for generating path explanations). Table \ref{tab:runtime_comparison} reports the concrete runtime of representative compared methods in three datasets. For the recommendation phase, PAGSL's training time is comparable to other GNN-based recommendation methods. The additional overhead introduced by the path-tree encoding is modest and is shared among queries during inference, making PAGSL suitable for large-scale deployment.
For the explanation phase, PAGSL requires 0.91–1.01 seconds per query across the three datasets. This efficiency advantage stems from the $k$-shell decomposition, which reduces the augmented graph before mask learning, substantially shrinking the search space for the subsequent optimization. 
It is worth noting that while PGExp-SR exhibits fast online inference speeds, it requires a substantial amount of offline pre-training time to learn the explanation network. In contrast, PAGSL avoids such heavy offline training overhead while still maintaining highly practical deployment speeds. 
This confirms that our multi-level pruning mechanism not only ensures explanation fidelity but also guarantees practical deployment speed, significantly outperforming traditional explanatory models.

\begin{table*}[!ht]
    \centering
    \begin{threeparttable}
    \caption{Empirical runtime comparison on three datasets.}
    \label{tab:runtime_comparison}
    
    \renewcommand{\arraystretch}{1.2}
    
    \small
    \setlength{\tabcolsep}{3.0pt}
        
        \begin{tabular*}{\textwidth}{@{\extracolsep{\fill}} >{}l *{6}{>{}c} }  
            \toprule
            \multirow{2}{*}{\textbf{Method}} & \multicolumn{2}{c}{\textbf{Scholat}} & \multicolumn{2}{c}{\textbf{AMiner}} & \multicolumn{2}{c}{\textbf{ResearchGate}} \\
            \cmidrule(lr){2-3} \cmidrule(lr){4-5} \cmidrule(lr){6-7}
            & Rec. & Explan. & Rec. & Explan.& Rec. & Explan. \\
            & (s / epoch) & (s / query) & (s / epoch) & (s / query) & (s / epoch) & (s / query) \\
            \midrule
            LightGCN & 3.7287 & -- & 3.1764 & -- & 0.8087 & -- \\
            SIGMA & 4.4331 & -- & 3.0027 & -- & 1.1166 & -- \\
            HGCL & 3.8303 & -- & 3.4254 & -- & 1.0835 & -- \\
            DiffKG & 4.3652 & -- & 3.5007 & -- & 1.9474 & -- \\
            
            \midrule
            GNNExp-SR & -- & 1.7142 & -- & 2.1613 & -- & 1.3253 \\
            PGExp-SR & -- & 0.0868 & -- & 0.1022 & -- & 0.0558 \\
    
            \midrule
            PAGSL & \textbf{4.0661} & \textbf{1.0100} & \textbf{3.3877} & \textbf{0.9443} & \textbf{0.9920} & \textbf{0.9110} \\
            \bottomrule
        \end{tabular*}
  \end{threeparttable}
\end{table*}

\subsection{Ablation Study}
\subsubsection{\textbf{Recommendation Performance}}
In this subsection, we design three ablation variants by selectively disabling key modules. The ablated configurations are described as follows:

\begin{itemize}[noitemsep, topsep=2pt, leftmargin=*]
    \item \textbf{w/o MRC}: This variant removes the multiple relation convolution (MRC) module responsible for encoding heterogeneous edge semantics. Scholar representations are derived solely from iterative path encoding, without explicit modeling of edge-type-specific interactions.
    
    \item \textbf{w/o PAE}: This configuration omits the path-aware encoding (PAE) module, thereby preventing the aggregation of multi-relational neighbor features along structured academic paths.

    \item \textbf{w/o CL}: This variant disables the InfoNCE-based  objective, eliminating the semantic alignment between different augmented views of scholar representations.
\end{itemize}

The ablation results are summarized in Table~\ref{tab:ablation_study}, the following conclusions can be drawn:

\begin{itemize}[leftmargin=*,topsep=2pt]
    \item[(i)] \textbf{Effect of Heterogeneous Convolution.} The exclusion of the MRC module leads to a marked decline in performance across all datasets and metrics. This highlights the necessity of incorporating relation-aware message passing, which enables the model to distinguish and integrate heterogeneous academic interactions effectively.

    \item[(ii)] \textbf{Effect of Path Representation Learning.} Removing the PAE module also results in notable performance degradation. This validates the role of path aggregation in enriching scholar representations. The effect is particularly evident on the Scholat dataset, which benefits from denser and semantically richer paths, suggesting that path-level modeling is especially valuable in such contexts.

    \item[(iii)] \textbf{Effect of Contrastive Learning.} The absence of the contrastive learning objective consistently reduces performance, albeit to a lesser extent. This observation confirms the complementary benefit of contrastive training, which enhances representational robustness by aligning semantic consistency across augmented graph views.
\end{itemize}

Taken together, these results demonstrate that each of the three components contributes a distinct and indispensable effect to the overall performance of PAGSL. The integration of heterogeneous relational learning, path-aware modeling, and contrastive objectives is crucial for achieving accurate and explainable scholar recommendation.

\begin{table}
    \centering
    \begin{threeparttable}
    \caption{Ablation study on key components of PAGSL for scholar recommendation tasks.}
    \label{tab:ablation_study}
        \renewcommand\arraystretch{1.5}
        \scriptsize
        \setlength{\tabcolsep}{3.0pt}
        \begin{tabular}{cccccccccccccc}
            \toprule 
            Dataset  & Models  & Precision & Recall & F1-score & ACC & AUC & HR@5 & NDCG@5 & HR@10 & NDCG@10 \\
            \midrule 
            \multirow{4}{*}{Scholat}
            & PAGSL w/o MRC & 0.8606 & 0.8799 & 0.8701 & 0.8687 & 0.9392 & 0.1996 & 0.0820 & 0.3493 & 0.1172  \\ 
            
            & PAGSL w/o PAE & 0.8638 & 0.8960 & 0.8796 & 0.8774 & 0.9420 & 0.2458 & 0.1094 & 0.3920 & 0.1443 \\
 
            & PAGSL w/o CL & 0.9337 & 0.9524 & 0.9430 & 0.9424 & 0.9798 & 0.2547 & 0.1112 & 0.4206 & 0.1521  \\ \cline{2-11} 

            & PAGSL(\texttt{cos}) & \textbf{0.9744} & \textbf{0.9696} & \textbf{0.9720} & \textbf{0.9721} & \textbf{0.9899} & \textbf{0.2711} & \textbf{0.1221} & \textbf{0.4457} & \textbf{0.1647} \\
            
            \midrule 

            \multirow{4}{*}{AMiner}
            & PAGSL w/o MRC & 0.8815 & 0.8697 & 0.8756 & 0.8764 & 0.9452 & 0.2736 & 0.1262 & 0.3655 & 0.1480 \\

            & PAGSL w/o PAE & 0.9534 & 0.9089 & 0.9306 & 0.9332 & 0.9768 &0.3107 & 0.1491 & 0.3947 & 0.1696 \\

            & PAGSL w/o CL & 0.9640 & 0.9242 & 0.9437 & 0.9449 & 0.9829 & 0.3371 & 0.1569 & 0.4209 & 0.1776 \\ \cline{2-11} 
            
            & PAGSL(\texttt{cos}) & \textbf{0.9874} & \textbf{0.9694} & \textbf{0.9783} & \textbf{0.9785} & \textbf{0.9856} & \textbf{0.3567} & \textbf{0.1703} & \textbf{0.4495} & \textbf{0.1935} \\
            \midrule 

            \multirow{4}{*}{ResearchGate}
            & PAGSL w/o MRC & 0.7519 & 0.7417 & 0.7468 & 0.7485 & 0.8189 & 0.1788 & 0.0544 & 0.3497 & 0.0771\\

            & PAGSL w/o PAE & 0.7423 & 0.8248 & 0.7814 & 0.7692 & 0.8491 & 0.2150 & 0.0688 & 0.4482 & 0.1027\\

            & PAGSL w/o CL & 0.7451 & \textbf{0.8717} & 0.8035 & 0.7868 & 0.8743 & 0.2202 & 0.0547 & 0.4275 & 0.0761 \\ 
            \cline{2-11} 

            & PAGSL(\texttt{cos}) & \textbf{0.8520} & 0.8457 & \textbf{0.8488} & \textbf{0.8494} & \textbf{0.9233} & \textbf{0.2358} & \textbf{0.0818} & \textbf{0.4637} & \textbf{0.1095}\\

            \bottomrule 
        \end{tabular}
    \end{threeparttable}
\end{table}

\subsubsection{Effort of hub avoidance on Explanation Quality}
To quantitatively evaluate the specificity of generated explanation paths and the effectiveness of hub avoidance, we compare our framework against two designed specific baselines:
(1) \textbf{Standard-SP}, which applies standard shortest path algorithm on the original augmented graph with same edge weights; and
(2) \textbf{Mask-Only-SP}, which utilizes the learned edge mask weights for path finding but entirely removes the node centrality penalty term from the score function. And we introduce two novel metrics and conduct a targeted ablation study. For the specificity of explanations and the model's ability to bypass hub nodes, we define the following two metrics:

\begin{itemize}
\item \textbf{Average Node Degree of Paths (AND-P)}: AND-P measures the average connectivity of the intermediate nodes constituting the explanation paths:
\begin{equation}
    \text{AND-P} = \frac{1}{\left| \Pi \right|}\sum_{p \in \Pi}\left( \frac{1}{\left| V_{int}^{\left(p\right)} \right|} \sum_{v\in V_{int}^{\left(p\right)}} \text{Degree}\left(v \right) \right)
\end{equation}
where $\Pi$ is the set of generated explanation paths, for each path $p \in \Pi$, $V_{int}^{(p)}$ denotes the set of its intermediate nodes. A significantly lower AND-P directly demonstrates the model's success in avoiding high-degree generic nodes.

\item \textbf{Hub Inclusion Ratio (HIR-$k$\%)}: HIR-$k$\% denotes the set of top $k$\% highest-degree nodes in the graph as the hub set $\mathcal{H}_k$ (we set $k=1$ for our evaluation):
\begin{equation}
    \text{HIR-k\%} = \frac{1}{\left| \Pi \right|}\sum_{p\in \Pi}\mathbb{I}\left( \exists v \in V_{int}^{\left(p\right)} \ \text{s.t.} \ v \in \mathcal{H}_k \right)
\end{equation}
where $\mathbb{I}(\cdot)$ is the indicator function. A lower HIR value indicates that the generated explanations do not rely on universally connected core nodes, thereby exhibiting higher specificity.
\end{itemize}

\begin{table*}[!ht]
    \centering
    \begin{threeparttable}
    \caption{Ablation study on key components of PAGSL for recommendation explanation tasks.}
    \label{tab:explananation_ablation_study}
    
    \renewcommand{\arraystretch}{1.3}
    
    \scriptsize
    \setlength{\tabcolsep}{3.0pt}
        
        \begin{tabular*}{\textwidth}{@{\extracolsep{\fill}} >{}l *{6}{>{}c} }
            \toprule
            \multirow{2}{*}{\textbf{Method}} & \multicolumn{2}{c}{\textbf{Scholat}} & \multicolumn{2}{c}{\textbf{AMiner}} & \multicolumn{2}{c}{\textbf{ResearchGate}} \\
            \cmidrule(lr){2-3} \cmidrule(lr){4-5} \cmidrule(lr){6-7}
            & AND-P & HIR-1\% & AND-P & HIR-1\% & AND-P & HIR-1\% \\
            \midrule
            Standard-SP & 158.8137 & 0.7502 & 98.7515 & 0.7929 & 75.8266 & 0.6548 \\
            Mask-Only-SP & 102.4866 & 0.3132 & 53.6567 & 0.4621 & 47.5012 & 0.3906 \\
    
            \midrule
            PAGSL & \textbf{74.3637} & \textbf{0.1032} & \textbf{27.6667} & \textbf{0.1585} & \textbf{37.5017} & \textbf{0.1906} \\
            \bottomrule
        \end{tabular*}
  \end{threeparttable}
\end{table*}

The quantitative results across the three datasets are summarized in Table \ref{tab:explananation_ablation_study}. 
As observed, our PAGSL method achieves substantially lower AND-P and HIR-1\% values compared to all baselines. The comparison between Standard-SP and Mask-SP demonstrates that edge mask learning alone provides some degree of hub avoidance by down-weighting uninformative edges. The further reduction from Mask-SP to our full method confirms the additional and significant contribution of the explicit centrality penalty in the score function $\mathcal{E}_p$. These results provide direct quantitative evidence that our framework generates explanation paths that are more specific and less reliant on generic hub nodes, aligning with our design motivation.

\subsubsection{Effect of $k$-shell Decomposition on Explanation Quality}
\label{sec:ablation_k-shell}
\begin{table}[htbp]
    \centering
    \begin{threeparttable}
    \caption{Effect of $k$-shell decomposition across datasets.}
    \label{tab:kshell_effect}
    \small
    \setlength{\tabcolsep}{3.0pt}
    \begin{tabular}{llccccc}
        \toprule
        Dataset & Variant & Fidelity & AUC & Node Red.(\%) & Edge Red.(\%) & Time(s/query) \\
        \midrule
        Scholat & w/o $k$-Shell & 0.0204 & 0.8497 & 0 & 0 & 1.2344 \\
         & \textbf{PAGSL} & \textbf{0.0177} & \textbf{0.8458} & \textbf{13.49\%} & \textbf{14.55\%} & \textbf{1.0100} \\
        \addlinespace
        AMiner & w/o $k$-Shell & 0.0767 & 0.8588 & 0 & 0 & 1.5100 \\
         & \textbf{PAGSL} & \textbf{0.0745} & \textbf{0.8571} & \textbf{34.36\%} & \textbf{46.70\%} & \textbf{0.9443} \\
        \addlinespace
        ResearchGate & w/o $k$-Shell & 0.1072 & 0.8022 & 0 & 0 & 1.3341 \\
         & \textbf{PAGSL} & \textbf{0.0831} & \textbf{0.8010} & \textbf{20.53\%} & \textbf{20.10\%} & \textbf{0.9110} \\
        \bottomrule
    \end{tabular}
    \end{threeparttable}
\end{table}

To validate the effect of $k$-shell decomposition on explanation fidelity, we conduct an ablation experiment by comparing explanation quality with and without $k$-shell decomposition, as shown in Table \ref{tab:kshell_effect}. We measure explanation fidelity using the Fidelity Score, defined as follows,
\begin{equation}
    \text{Fidelity} = \frac{1}{|V|} \sum_{(v_i, v_j) \in V} \left[ f(G, v_i, v_j) - f(G \setminus \Pi_{ij}, v_i, v_j) \right]
\end{equation}
where $f(G, v_i, v_j)$ is the model's prediction score for the pair $(v_i, v_j)$ on graph $G$, $\Pi_{ij}$ is the set of explanation paths. Fidelity denotes the average drop in prediction confidence when explanation path edges are removed from the graph. Node Red. and Edge Red. represent the percentage of nodes and edges removed by $k$-shell decomposition, respectively. As observed, the AUC values of two variants are nearly identical across all datasets, indicating that $k$-shell decomposition causes negligible loss in edge-level explanation accuracy. The Fidelity differences are small in absolute terms, confirming that the core explanatory structure is well preserved after pruning. 
Furthermore, $k$-shell decomposition achieves substantial graph simplification (13.49–34.36\% node reduction and 14.55–46.70\% edge reduction). This demonstrates that a significant portion of the augmented graph consists of peripheral nodes that do not participate in any valid reasoning path. Consequently, PAGSL consistently requires less time per explanation query than the w/o $k$-Shell variant across all datasets.
These results collectively demonstrate that $k$-shell decomposition is not merely a computational speedup but a quality-preserving structural refinement: it removes noise introduced by the augmentation process without sacrificing any semantically meaningful explanation path.

\subsection{\textbf{Parameter Analysis}}

\begin{enumerate}[label=(\arabic*), leftmargin=*]
    \item \textbf{Recommendation Performance}
\end{enumerate}

\begin{figure}
     \centering
        \includegraphics[width=\linewidth]{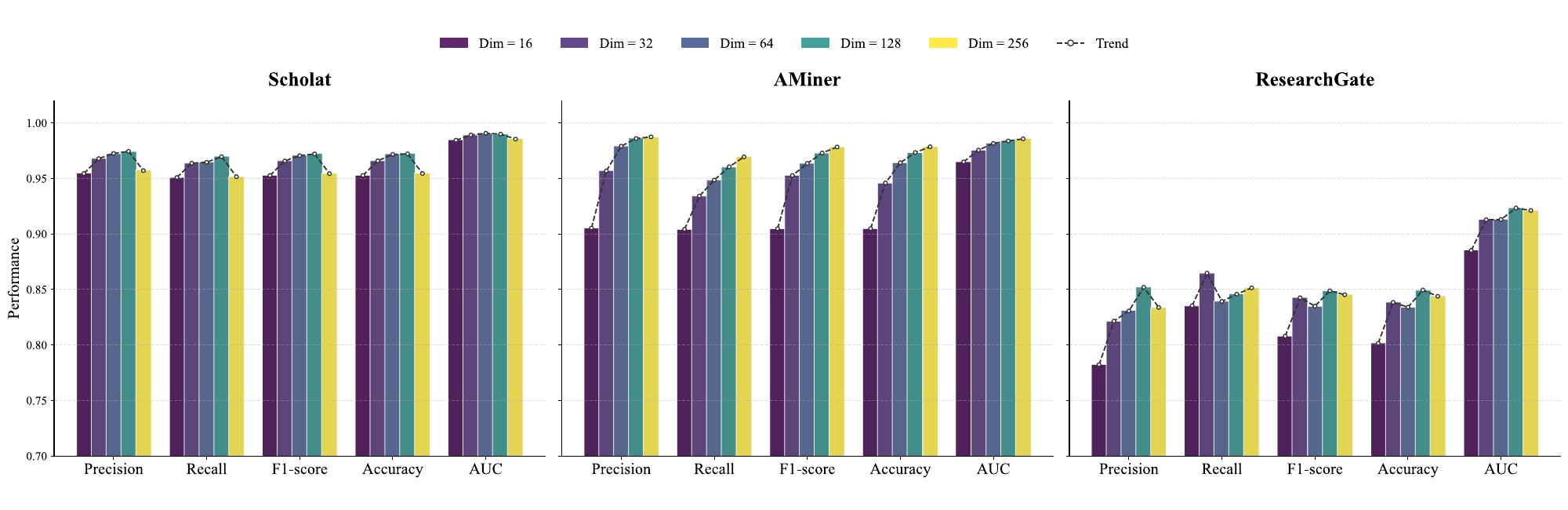}
    \caption{The experimental results of different embeddings dimension on three datasets with classification metrics.}
    \label{fig:dimension_class}
\end{figure}

\begin{figure}
     \centering
        \includegraphics[width=\linewidth]{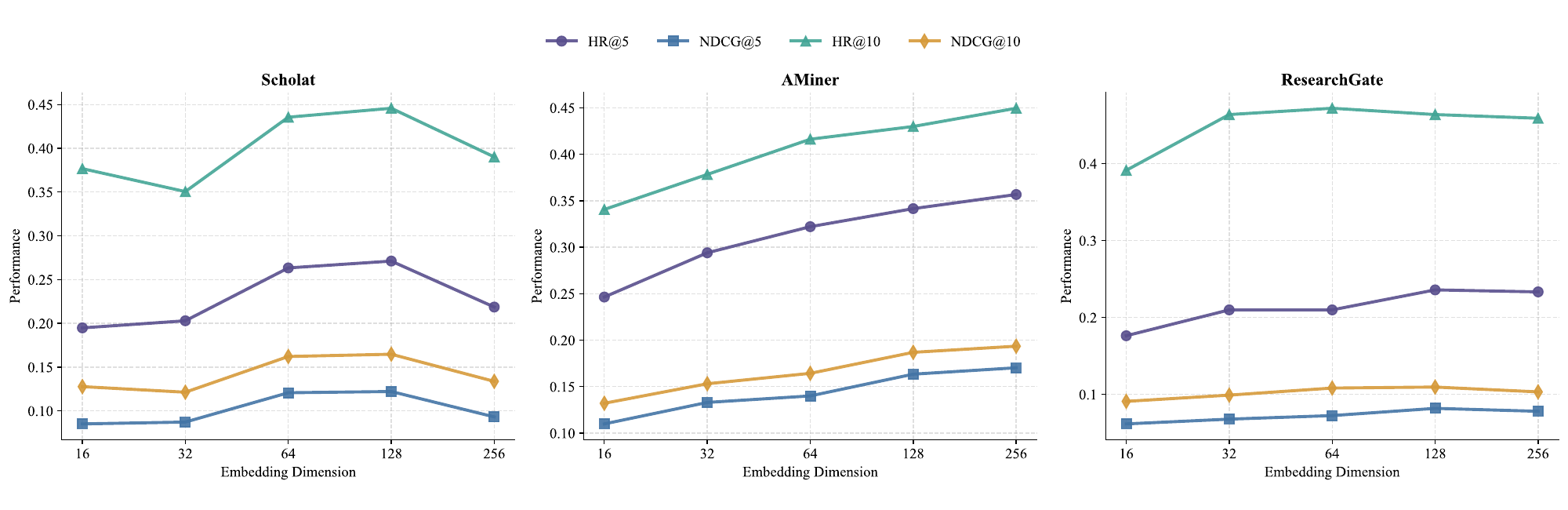}
    \caption{The experimental results of different embeddings dimension on three datasets with ranking metrics.}
    \label{fig:dimension_rank}
\end{figure}

\begin{figure}
     \centering
        \includegraphics[width=\linewidth]{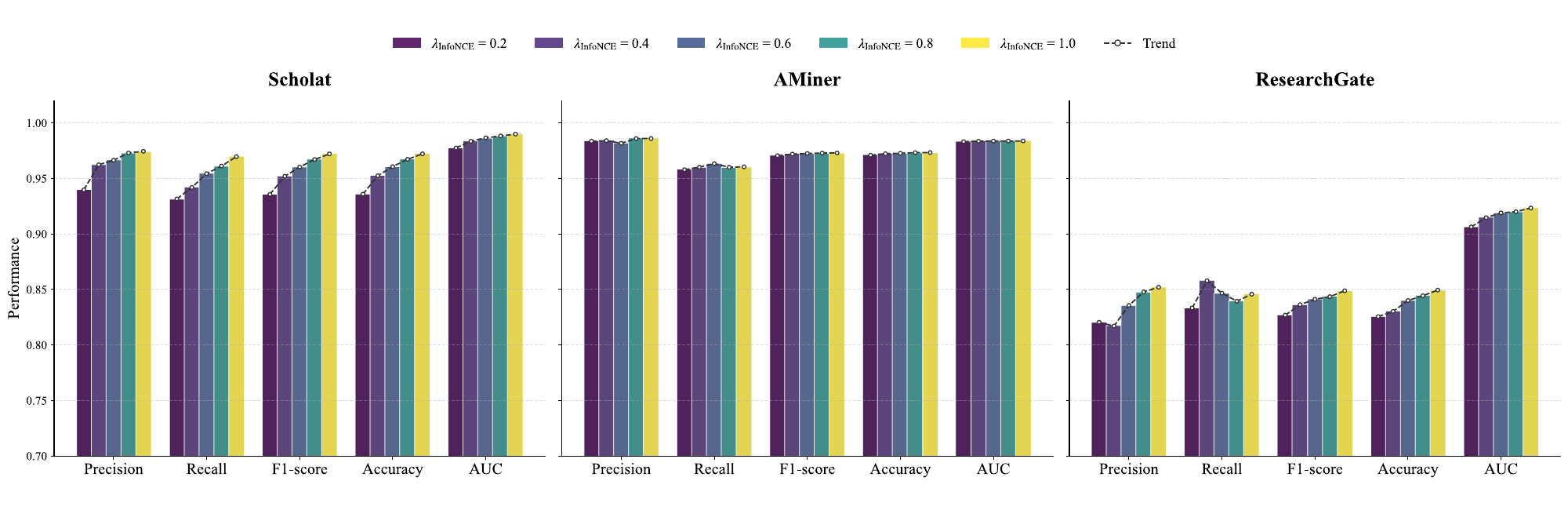}
    \caption{The experimental results of different contrastive loss coefficient $\lambda_\text{InfoNCE}$ on three datasets with classification metrics.}
    \label{fig:infonce_class}
\end{figure}

\begin{figure}
     \centering
        \includegraphics[width=\linewidth]{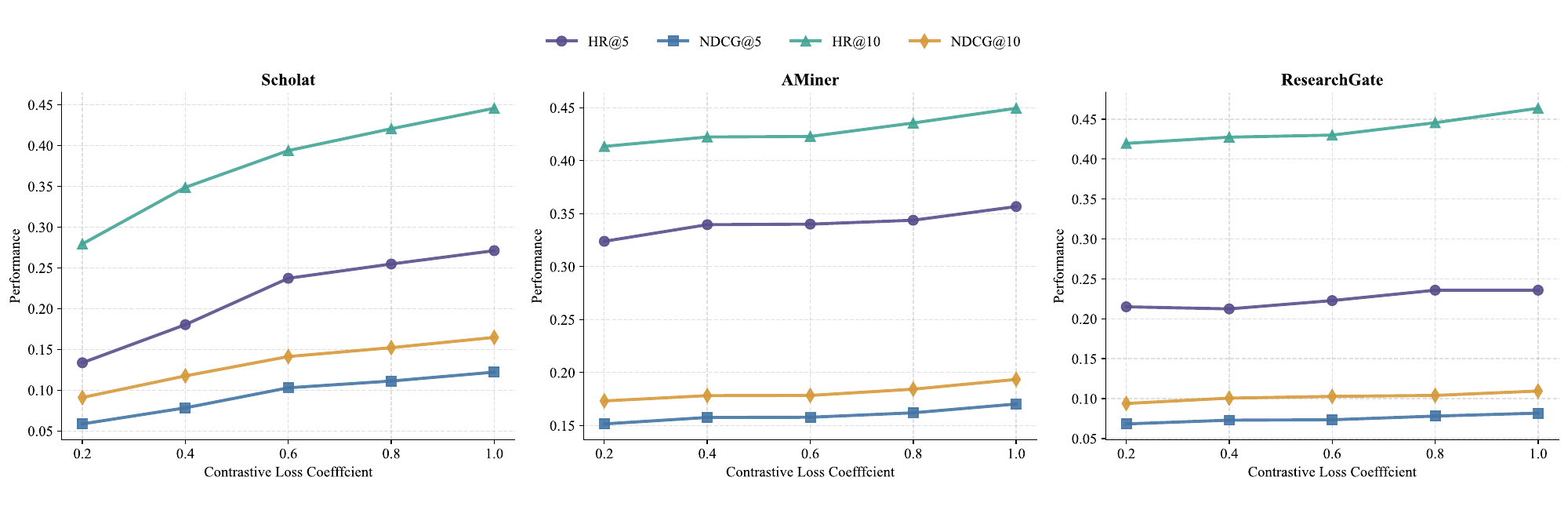}
    \caption{The experimental results of different contrastive loss coefficient $\lambda_\text{InfoNCE}$ on three datasets with ranking metrics.}
    \label{fig:infonce_rank}
\end{figure}

In the scholar recommendation task, PAGSL encodes multiple academic behavioral paths through a path-aware gnn, and applies a prediction head to estimate scholar relevance scores. We investigate the impact of several important hyperparameters, based on the best-performing prediction head identified for each dataset—namely, \texttt{cos} for Scholat, \texttt{dot} for AMiner, and \texttt{mlp} for ResearchGate. The following parameters are studied in detail:

\begin{itemize}[noitemsep, topsep=2pt, leftmargin=*]
    \item \textbf{Embedding Dimension.} This parameter determines the dimensionality of the node representation vectors. We vary the embedding size in $\{16, 32, 64, 128, 256\}$ and report the results in Figure~\ref{fig:dimension_class} and Figure~\ref{fig:dimension_rank}. The optimal dimension is selected as 128 for Scholat and ResearchGate, and 256 for AMiner, indicating that higher-dimensional embeddings are more beneficial under sparse interaction settings.

    \item \textbf{Contrastive Loss Coefficient.} This coefficient controls the relative importance of the InfoNCE-based contrastive loss during model optimization. We evaluate values from $\{0.2, 0.4, 0.6, 0.8, 1.0\}$, as shown in Figure~\ref{fig:infonce_class} and Figure~\ref{fig:infonce_rank}. The best performance across all datasets is consistently achieved when the coefficient is set to 1.0, suggesting that contrastive supervision plays a critical role in enhancing representation robustness.
\end{itemize}

\begin{enumerate}[label=(\arabic*), start=2, leftmargin=*]
    \item \textbf{Explanation Quality}
\end{enumerate}

\begin{figure}
     \centering
        \includegraphics[scale=.8]{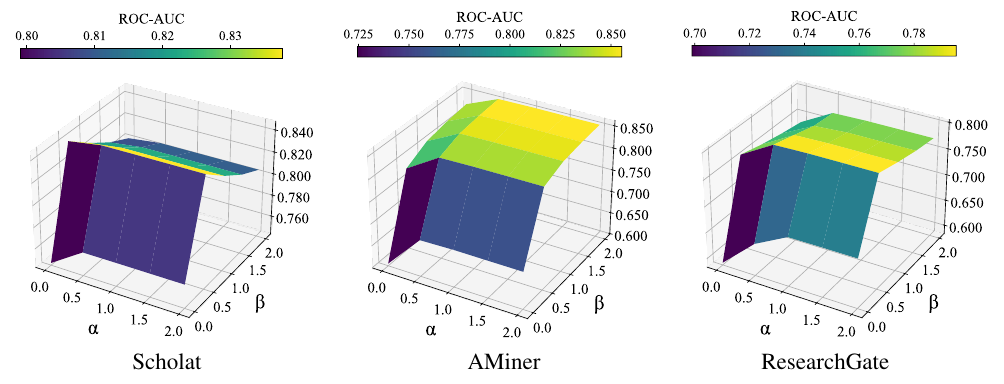}
    \caption{The impact of hyperparameters $\alpha,\beta$ on three datasets.}
    \label{fig:alpha_beta}
\end{figure}

For explainability evaluation, PAGSL utilizes an edge mask optimization mechanism to identify semantically informative paths for each recommendation instance. The mask weights are dynamically updated based on the activation of specific scholars. We study the sensitivity of explanation quality with respect to the following parameters:

\begin{itemize}[noitemsep, topsep=2pt, leftmargin=*]
    \item \boldmath$\alpha$ and $\beta$\unboldmath. These two coefficients control the magnitude of regularization applied to activated (on-path) and inactivated (off-path) edges, respectively, during mask optimization. We experiment with $\{0.1, 0.5, 1.0, 1.5, 2.0\}$ for both parameters, and report the results in Figure~\ref{fig:alpha_beta}. The optimal settings are found to be $(\alpha=1.0, \beta=0.5)$ for Scholat, $(\alpha=1.0, \beta=2.0)$ for AMiner, and $(\alpha=1.5, \beta=0.5)$ for ResearchGate. These results suggest that a moderate penalty on irrelevant edges and a relatively higher reward for important ones contribute to more precise and interpretable path selection.
\end{itemize}

\section{Discusssion}
\subsection{Theoretical implications}
Our proposed PAGSL integrates augmented relation learning and topological path masking to resolve three fundamental challenges in explainable scholar recommendation: semantic sparsity, non-linear academic influence propagation, and topological bias in explanation.To address the limitations of traditional structure-dependent paradigms, we redefine the boundaries of GSL via an attribute-driven augmentation strategy, explicitly encoding the homophily hypothesis to establish semantic bridges for cold-start and long-tail nodes. Furthermore, recognizing that sequential meta-paths fail to capture the branching nature of academic influence, we design a path-tree heterogeneous learning module that expands linear interaction chains into hierarchical trees for adaptive semantic aggregation. Finally, to counteract the hub node bias induced by the rich-club phenomenon in scale-free networks, we incorporate a hierarchically-guided path mask mechanism, theoretically positing that effective explanatory paths must traverse the network's core semantic layer rather than merely following degree centrality.

Crucially, the ASNs are usually sparse and may have noisy edges in real world. One concern is whether such sparse and noise effects hurt explanation generation. We found that our framework employs graph augmentation and path mask learning with centrality penalty to form a robust, multi-layered defense mechanism. 
Specifically, by establishing heterogeneous attribute edges between scholars, our attribute-shared graph augmentation provides abundant structural priors, enabling effective message passing and path-based explanation even for low-degree scholars. 
Afterwords,  $k$-shell decomposition effectively filters out structural noise by iteratively pruning low-degree nodes, as this noise lacks dense local clustering support, thereby achieving noise purification before mask optimization. 
For deeper semantic noise, the prediction-sensitive mask learning penalizes edges lacking informative signals, and the centrality penalty actively bypasses generic super hub nodes to ensure high-specificity explanations. These modules grant our model exceptional robustness in handling sparse graphs or noisy edges.

\subsection{Practical implications}
By integrating structural enhancement, heterogeneous modeling, and topology-aware denoising, PAGSL significantly advances the practical implementation of explainable scholar recommendation. It effectively mitigates information overload and addresses the "cold-start" problem for early-career researchers by identifying latent collaboration opportunities based on intrinsic interests, fostering a fairer allocation of academic resources. Furthermore, its hierarchically-aware explanations provide critical support for decision-making, such as reviewer assignment and funding panel selection. PAGSL produces transparent rationales that align with expert cognitive reasoning. This reduces cognitive load, facilitates trust calibration, and enhances user engagement. Finally, PAGSL's robustness against structural noise and hub-node bias renders it highly applicable to large-scale and dynamic networks. This extends its utility to tasks like trend prediction and team formation,  providing a robust foundation for trustworthy academic intelligence platforms.

While this paper focuses on scholar recommendation in ASNs, the proposed framework possesses a degree of generality that enables adaptation to other domains. This generalizability stems from two architectural properties. First, the explanation module operates post-hoc on the trained model without modifying its structure or parameters, making it an instance-level explainer that is decoupled from the specific recommendation task. Second, explanation paths are extracted from the augmented graph's typed edges rather than from domain-specific relational schema, avoiding the need for expert-designed meta-paths in each new domain.
To apply the framework in a new domain (such as e-commerce recommendations), the primary adaptation involves defining domain-specific node attributes and attribute-sharing rules for graph augmentation. The explanation module requires no structural modification across domains, as it operates on the general abstractions of typed edges and node centrality. The hyperparameter controlling the centrality penalty may require domain-specific tuning to adapt to the unique degree distributions of the target network. This adaptability and instance-level flexibility make PAGSL a highly versatile framework for providing transparent and faithful explanations across diverse complex network applications.

\section{Conclusions and Future Work}
\label{sec:conclusion}
This paper introduces a path-augmented graph structure learning framework for explainable scholar recommendation in academic social networks. PAGSL addresses two core challenges: achieving accurate scholar matching and providing path-based explanations.
It consists of four main steps. First, textual attribute information of scholars is processed to construct attribute-based relationship paths, augmenting the original user-following relationship network. Then, a graph relation learning model based a path neural network is developed to learn the embedding representation of each scholar, and the appropriate prediction head is selected for scholar recommendation based on the embedding results. During the explanation of recommendation results, we utilize $k$-shell decomposition to identify multiple paths of nodes, and subsequently apply mask learning to heterogeneous paths among scholars to determine critical explanatory paths. By processing heterogeneity and explaining edges, the method identifies significant paths linking source nodes to target nodes.
Extensive experiments across three benchmark datasets demonstrate that PAGSL consistently outperforms strong baselines, with improvements ranging from 3\% to 26\% across key metrics.

Moreover, both quantitative and qualitative analyses confirm the model’s ability to deliver informative explanations grounded in collaborative behavior and attribute semantics.
Future work will involve optimizing path selection algorithms~\citep{li2021graphmse} to reduce path quantity and complexity. 
In addition, multi-objective optimization techniques will be explored to determine the optimal balance between explainability and predictive accuracy, thereby maximizing interpretability without sacrificing predictive performance.

\newpage


\bibliographystyle{model5-names}
\bibliography{main}


\end{document}